\documentclass[twocolumn,english,aps,pra,showpacs,amsfonts,amsmath,amssymb,superscriptaddress,footinbib,floatfix,longbibliography]{revtex4-1}
\usepackage[T1]{fontenc}
\usepackage[latin9]{inputenc}
\usepackage{mathtools}
\usepackage{amsmath}
\usepackage{amssymb}
\usepackage{graphicx}

\makeatletter
\usepackage{ulem}
\usepackage{color}
\def\kbar{{\mathchar'26\mkern-9mu k}}
\usepackage{filemod}

\newcommand{\hidetxt}[1]{}

\usepackage{babel}

\makeatother

\usepackage{babel}
\begin{document}

\title{Experimental realization of an ideal Floquet disordered system}

\author{Cl\'ement Hainaut}

\affiliation{Universit\'e de Lille, CNRS, UMR 8523 -- PhLAM -- Laboratoire
de Physique des Lasers Atomes et Mol\'ecules, F-59000 Lille, France}

\author{Adam Ran\c con}

\affiliation{Universit\'e de Lille, CNRS, UMR 8523 -- PhLAM -- Laboratoire
de Physique des Lasers Atomes et Mol\'ecules, F-59000 Lille, France}

\author{Jean-Fran\c cois Cl\'ement]}
\affiliation{Universit\'e de Lille, CNRS, UMR 8523 -- PhLAM -- Laboratoire
de Physique des Lasers Atomes et Mol\'ecules, F-59000 Lille, France}

\author{Isam Manai}

\affiliation{Universit\'e de Lille, CNRS, UMR 8523 -- PhLAM -- Laboratoire
de Physique des Lasers Atomes et Mol\'ecules, F-59000 Lille, France}

\author{Pascal Szriftgiser}

\affiliation{Universit\'e de Lille, CNRS, UMR 8523 -- PhLAM -- Laboratoire
de Physique des Lasers Atomes et Mol\'ecules, F-59000 Lille, France}

\author{Dominique Delande}

\affiliation{Laboratoire Kastler Brossel, UPMC, CNRS, ENS, Coll{\`e}ge de France;
4 Place Jussieu, F-75005 Paris, France}

\author{Jean Claude Garreau}

\affiliation{Universit\'e de Lille, CNRS, UMR 8523 -- PhLAM -- Laboratoire
de Physique des Lasers Atomes et Mol\'ecules, F-59000 Lille, France}

\author{Radu Chicireanu}

\affiliation{Universit\'e de Lille, CNRS, UMR 8523 -- PhLAM -- Laboratoire
de Physique des Lasers Atomes et Mol\'ecules, F-59000 Lille, France}
\begin{abstract}
The atomic Quantum Kicked Rotor  is an outstanding ``quantum
simulator'' for the exploration of transport in disordered quantum systems.
Here we study experimentally the \emph{phase-shifted
quantum kicked rotor}, which we show to display properties close to an ideal
disordered quantum system, opening new windows into the study of Anderson
physics. 
\end{abstract}

\date{\filemodprint{\jobname}~~File: \jobname}

\maketitle

\section{Introduction}

Transport phenomena are omnipresent in physics. Hydrodynamics (i.e.
matter transport), heat (i.e. energy transport) and electrical conductivity
(i.e. charge transport) are among the most practically important examples.
Less mundane microscopic examples are coherence or spin polarization transport.
The common feature of the (macroscopic) classical approach to these
phenomena is a coarse grain average leading to partial differential
equations valid if the typical sizes of the system are much larger
than those of its individual parts, an approach which tends to wane
microscopic features. However, when one considers mesoscopic systems,
at some scale the microscopic structure come into play, and quantum
phenomena (notably quantum interferences) should be considered. This
is the case when one treats e.g. superfluid helium, or electron transport
in small enough structures (e.g. quantum dots) or  motion of ultracold atoms
in optical lattices. In such cases, quantum dynamics often becomes
dominant and the classical transport equations break down, due to
new phenomena like superconductivity, superfluidity, quantum phase
transitions, etc. 

A very important issue is quantum (or wave)
transport in disordered systems, which has lead for more than 60 years
to a wealth of theoretical and experimental studies, starting with
the celebrated Anderson model~\cite{Anderson:LocAnderson:PR58} describing,
in a relatively simple (and tractable) way, the ``localization'' of the
electron wave function in a disordered crystal. Recently,  the concept of ``quantum
simulation'' of complicated quantum systems has been introduced~\cite{Mazza:OpticalLatticeBasedQuantumSimulator:NJP12,Bloch:QuantumSimulationsUltracoldGases:NP14,
Georgescu:QuantumSimulation:RMP14}. The
main idea, inspired from early insights by Feynman~\cite{Feynman:SimulatingPhysics:IJTP82},
is to engineer a (relatively) simple, controlled system able to reproduce
(some of) the quantum features of a complex, less controllable one.
A main conducting line in this field is to simulate condensed matter
systems using ultracold atoms in optical lattices~\cite{Bloch:ManyBodyUltracold:RMP08},
which allowed for perfectly controlled realizations of the Hubbard hamiltonian, the observation of the Mott transition~\cite{Greiner:MottTransition:N02,Paredes:TonksGirardeauGas:N04},
or the observation of  Anderson localization in disordered systems~\cite{Moore:AtomOpticsRealizationQKR:PRL95,Billy:AndersonBEC1D:N08,Roati:AubryAndreBEC1D:N08,
Jendrzejewski:AndersonLoc3D:NP12,Semeghini:MobilityEdgeAnderson:NP15}. 
A particularly useful quantum simulator for disordered transport is
the so-called atomic quantum kicked rotor (QKR), first realized by Raizen and co-workers~\cite{Moore:AtomOpticsRealizationQKR:PRL95},
and developed by our group~\cite{Chabe:Anderson:PRL08,Lopez:ExperimentalTestOfUniversality:PRL12,Manai:Anderson2DKR:PRL15,
Hainaut:CFS:NCM18,Garreau:QuantumSimulationOfDisordered:CRP17}.

While the QKR has proved to be an excellent quantum simulator of Anderson localization, it displays temporal correlations between the kicks, which are equivalent to a spatially correlated disorder in the Anderson model.
In the present work we show that by engineering the kicked rotor's Hamiltonian one
can mimic a nearly-uncorrelated disordered system, allowing a much
more precise comparison between theory and experiment. This is done
by introducing  periodical phase shifts to the kicking potential,
giving rise to the  \emph{periodically-shifted quantum kicked
rotor} (PSQKR), which allows for a very efficient averaging over disorder
realizations, able to erase the undesired effects of kicks correlations.
An interesting study of the interplay between classical and quantum
transport in the PSQKR can be found in Ref.~\cite{Hainaut:RatchetEffectQKR:PRA18}.
We will benchmark our experimental PSQKR results by comparison with a numerical variant of the QKR, the Random Kicked Rotor (RKR), which mimics a perfectly uncorrelated disorder in the Anderson model, but which is not realizable experimentally.
The power of the PSQKR is evidenced by measuring,
with unprecedented precision, the universal transport properties predicted
by the one-parameter scaling theory~\cite{Abrahams:Scaling:PRL79}. 

The article is organized as follows. In Section~\ref{sec_QKR}, we briefly review the QKR and its transport properties, as well as its connection to the Anderson model.
Section~\ref{sec:PSQKR} describes
the model and the experimental realization of the PSQKR.
Section~\ref{sec:Dini} compares the diffusion coefficient obtained in
that way with the one extracted from numerical simulations of the RKR, demonstrating
the ability of the PSQKR to average out the temporal correlations. Section~\ref{sec_beta}
describes the experimental measurement of the universal scaling
function $\beta(g)$, evidencing the excellent agreement between the PSQKR
and the numerical predictions using the RKR, as well
as with the weak-localization theoretical predictions. Section~\ref{sec_concl}
concludes the work.

\section{Quantum transport in the atomic kicked rotor \label{sec_QKR}}

The atomic kicked rotor first demonstrated by Raizen's group consists of a cloud of laser-cooled atoms submitted to a periodically-pulsed laser standing wave (SW)~\cite{Moore:AtomOpticsRealizationQKR:PRL95}. This first breakthrough led to an impressive corpus of experimental results in different fields as
e.g. quantum transport~\cite{Philips:QuantumResTalbot:PRL99,Summy:QRRarchets:PRL08,Rubins-Dunlop:ChaosAssitTunnel:Nature01,Raizen:ChaosAssistTunnel:Science01,Leonhardt:TransportKR:PRL05,Wimberger:QRScalingLaw:PRA05} and quantum metrology~\cite{DArcy:GravityQuantRes:PRL04}. In quite general conditions, for short enough times ($t\ll t_{\mathrm{loc}}$ where $t_{\mathrm{loc}}$ is the so-called ``localization time'') the atoms diffuse in momentum space with a second moment increasing linearly with time, $\left\langle p^{2}(t)\right\rangle =\left\langle p_{0}^{2}\right\rangle +2Dt$, but for $t\gtrsim t_{\mathrm{loc}}$, once quantum interferences build up in the system, the second moment saturates at a constant value $\left\langle p_{0}^{2}\right\rangle+p_{\mathrm{loc}}^{2}$. Here, $p_{\mathrm{loc}}$ is the localization ``length'' in momentum space, and $\left\langle p_{0}^{2}\right\rangle$ is the momentum variance of the initial wavepacket. It turns out that this phenomenon, called ``dynamical localization''~\cite{Casati:LocDynFirst:LNP79} \textendash{} i.e. localization in momentum space \textendash , is an analog of the (spatial) Anderson localization~\cite{Fishman:LocDynAnderson:PRA84} observed in disordered systems~\cite{Anderson:LocAnderson:PR58}. Without loss of generality, we will consider in the following only the case of narrow ($\left\langle p_{0}^{2}\right\rangle \ll p_{\mathrm{loc}}^{2}$) initial wavepakets, localized around $\left\langle p(0)\right\rangle=0$ (see experimental details below), and put $\left\langle p_{0}^{2}\right\rangle\simeq 0$, unless stated otherwise.

The kicked rotor's Hamiltonian is given by
\begin{equation}
H_{\textrm{QKR}}=\frac{p^{2}}{2}+K\sum_{k=0}^{\infty}\cos x\;\delta\left(t-k\right),\label{eq:QKR}
\end{equation}
where $K\cos x$ represents the sinusoidal potential created by the
standing wave (formed by counterpropagating laser beams of wave number $k_{L}$),
with the dimensionless spatial variable $x$ measured in units of
$\left(2k_{L}\right)^{-1}$ and the dimensionless time measured in
units of the kick period $T_{1}$. With these definitions $x$ and
$p$ obey the canonical commutation relation $\left[x,p\right]=i\kbar$
with $\kbar=4\hbar k_{L}^{2}T_{1}/M$ ($M$ the mass of the atom)
playing the role of a reduced Planck constant. The ``stochasticity parameter'' $K$ is given by $K=\kbar\Omega^{2}\tau/8
|\Delta|$, where $\Omega$ is the single-beam resonant Rabi frequency, $\Delta$ the laser-atom detuning and
$\tau$ the duration of the standing wave pulses. $K$ and $\kbar$ can
be tuned in the experiment (see below). The fact that the potential
is spatially periodic means that the quasimomentum $q$ is a constant
of motion, i.e. a given momentum $p_{0}$ is coupled only to momenta
of the form $p_{0}+\ell\kbar$ with $\ell\in\mathbb{Z}$;
one can thus always reduce the concerned momenta to the form $(q+\ell)\kbar$
with $\ell\in\mathbb{Z}$ and $q\in(-1/2,1/2]$.

The Anderson model~\cite{Anderson:LocAnderson:PR58} is described by
a tight-binding Hamiltonian of the form
\begin{equation}
H_{A}=\sum_i\epsilon_{i}\left|u_{i}\right\rangle \left\langle u_{i}\right|+\sum_{i,j\neq0}t_{j}\left|u_{i}\right\rangle \left\langle u_{i+j}\right|,\label{eq:HA}
\end{equation}
where $\left|u_{i}\right\rangle$ are the Wannier states localized on the lattice site $i$. In the Anderson model, $\epsilon_{i}$ are random on-site energies
distributed in a box $\left[-W/2,W/2\right]$, and $t_j$ are the hopping amplitudes. For the kicked rotor,
one starts from the evolution operator over one period (Floquet
operator)
\begin{equation}
U(1)=\exp\left(-i\frac{(\hat{\ell}+q)^{2}}{2}\kbar\right)\exp\left(-i\frac{K}{\kbar}\cos\hat{x}\right)\label{eq:U(1)}
\end{equation}
where $\hat{\ell}\left|\ell\right\rangle =\ell\left|\ell\right\rangle$ acts on the state space of the strictly periodic system. 
The leftmost operator
in Eq.~(\ref{eq:U(1)}) corresponds to the free propagation between
kicks and the rightmost corresponds to the kick~\footnote{The kinetic energy term can be neglected compared to the $\delta$
function in the kick term.}. Fishman and co-workers~\cite{Fishman:LocDynAnders:PRL82,Fishman:LocDynAnderson:PRA84}
projected the eigenvalue equation $U(1)\left|v_{\omega}\right\rangle =\exp(-i\omega)\left|v_{\omega}\right\rangle $
defining the Floquet quasi-eigenstates on a quasimomentum family and, using
algebraic operator identities, could show that this operator maps
onto a tight-binding Hamiltonian of the form~(\ref{eq:HA}), with
the ``on-site energies'' $\epsilon_{i} \rightarrow \epsilon_{\ell}=\tan\left(\omega/2-\kbar{(\ell+q)}^{2}/4\right)$
and the ``hoppings'' $t_{j} \rightarrow t_{r}=(2\pi)^{-1}\int dxe^{-irx}\tan\left(K\cos x/2\kbar\right)$.
The on-sites energies, in contrast to Anderson's model, are perfectly
deterministic, but if $\kbar$ is incommensurate with $\pi$ they oscillate, creating a so-called ``pseudo-disorder''. The fact that
the kicked rotor indeed displays the same behavior as the Anderson
model has been confirmed by a large number of experimental and numerical
works~\footnote{See e.g. \protect\cite{Garreau:QuantumSimulationOfDisordered:CRP17} and references therein.}.

This pseudo-disorder is however correlated, because the deterministic
phase acquired during the free evolution following a given kick is correlated to the phase acquired before the kick. Although this effect tends to
be averaged as the number of kicks increase, it leads to deterministic
effects visible in the dynamics, for instance oscillations in the
diffusion coefficient as a function of $K$ and of $\kbar$~\cite{Shepelyansky:Bicolor:PD87},
as well as in oscillations of $\left\langle p^{2}(t)\right\rangle $. This becomes particularly critical e.g.~if one tries to reconstruct
the function $\beta$ describing the asymptotic behavior of the system~\cite{Abrahams:Scaling:PRL79},
which is given, for the QKR, by
\begin{equation}
\beta\equiv\frac{\partial\ln g}{\partial\ln L}=\frac{\partial\ln\left(\left\langle p^{2}(t)\right\rangle ^{1/2}/t\right)}{\partial\ln\left(\left\langle p^{2}(t)\right\rangle ^{1/2}\right)},\label{eq:betaQKR}
\end{equation}
with the appropriate definition of the ``dimensionless conductance''
being $g\equiv\left\langle p^{2}(t)\right\rangle ^{1/2}/t$ and $L\equiv\left\langle p^{2}(t)\right\rangle ^{1/2}$
a measure of the ``size'' of the system~\cite{Cherroret:AndersonNonlinearInteractions:PRL14}. It expresses the behavior of the conductance with the size of the system, $g\sim L^{\beta}$ and is thus the kicked rotor's analog of the $\beta$ one-parameter
scaling function introduced by Anderson and coworkers~\cite{Abrahams:Scaling:PRL79}. A crucial assumption of the scaling theory is that $\beta$ can be 
expressed as a function of only $g$ itself, so that $\beta(g)$ is a universal
function which governs the transport properties of the system. Being
a logarithmic derivative, this function is particularly sensitive to the oscillations of $\left\langle p^{2}(t)\right\rangle $,
which might mask its universal behavior (i.e. independent of the systems' microscopic details). 

In order to obtain the equivalent of a perfectly uncorrelated disorder, as idealized in
the Anderson model, the free propagation phases should be a completely
random set of i.i.d. random variables $\phi_\ell$ in $[0,2\pi)$ that is, one replaces the free propagation operator in Eq.~\eqref{eq:U(1)} by $\sum_\ell \exp(-i\phi_\ell) |\ell\rangle\langle\ell|$, the kicking
part of the Hamiltonian being the same. This model is called the Random Kicked Rotor (RKR), for which the correlations indeed vanish, but it is not easily feasible experimentally~\footnote{In the RKR, $\kbar$ is not defined, as the free propagation phases are random, the pertinent parameter that we call $K$ for simplicity is equivalent to $K/\kbar$ for the standard QKR.}.

We will show that the PSQKR can mimic a nearly-uncorrelated disordered system, thus allowing a precise comparison between theory and experiment. The SW phase-shifting leads to a very efficient averaging over disorder realizations, which allows canceling the effects of temporal correlations and reveals, with unprecedented accuracy, the universal transport properties predicted by the one-parameter scaling hypothesis.

\section{The periodic Phase-Shifted Quantum Kicked Rotor \label{sec:PSQKR}}

\subsection{Model}

The PSQKR is a modified
version of the standard QKR, Eq.~\eqref{eq:QKR}, in which
the spatial position of the kicking potential is periodically shifted,
leading to the Hamiltonian
\begin{equation}
H_{\textrm{PSQKR}}=\frac{p^{2}}{2}+K\sum_{k=0}^{\infty}\sum_{j=1}^{N}\cos\left(x-a_{j}\right)\delta\left(t-kN-j\right)\label{eq:PSQKR}
\end{equation}
where $\{a_{j}\}$ is an arbitrary sequence of $N$ numbers in $[0,2\pi)$,
repeated each $NT_{1}$ periods of the kicking (period $N$ in dimensionless units). If all $a_{j}\equiv0$ we retrieve the standard QKR.   The preservation of the time-periodicity means that it displays Floquet eigenstates and that essentially the same mapping to an Anderson-like Hamiltonian can be applied; dynamical localization is expected to appear with an enlarged time scale multiplied by $N$, which was indeed verified theoretically, numerically and experimentally, as shown in Ref.~\cite{Hainaut:CFS:NCM18} for a slightly more complex Hamiltonian with an additional periodic modulation of the kick intensity. Even more interestingly, one can show that the symmetry properties of the phase sequence $\{a_{j}\}$ control the parity ($P$) and time-reversal ($T$) symmetry of the Hamiltonian Eq.~(\ref{eq:PSQKR}): if the sequence is antisymmetric around an arbitrary axis (e.g. $N=3$ and $a_{j}=-a,0,a$, with the anti-symmetry axis at kick $j=2$) $PT$-symmetry is preserved and the system is in the so-called ``orthogonal'' universality class, otherwise this symmetry is broken and the system is in the ``unitary'' universality class~\cite{Hainaut:CFS:NCM18}. Symmetry breaking has a deep influence on the transport and localization properties of the system, which can be most clearly put into evidence by a measurement of the universal $\beta(g)$ function~\cite{Hainaut:CFS:NCM18}. The central point of our approach is that averaging many different realizations of the phase sequences $\{a_{j}\}$ (chosen so that they break or not the $PT$-symmetry) is equivalent to a very efficient averaging over disorder realizations in the equivalent Anderson model.

\begin{figure*}[t!]
	\centering{}\centering \includegraphics[width=1\linewidth]{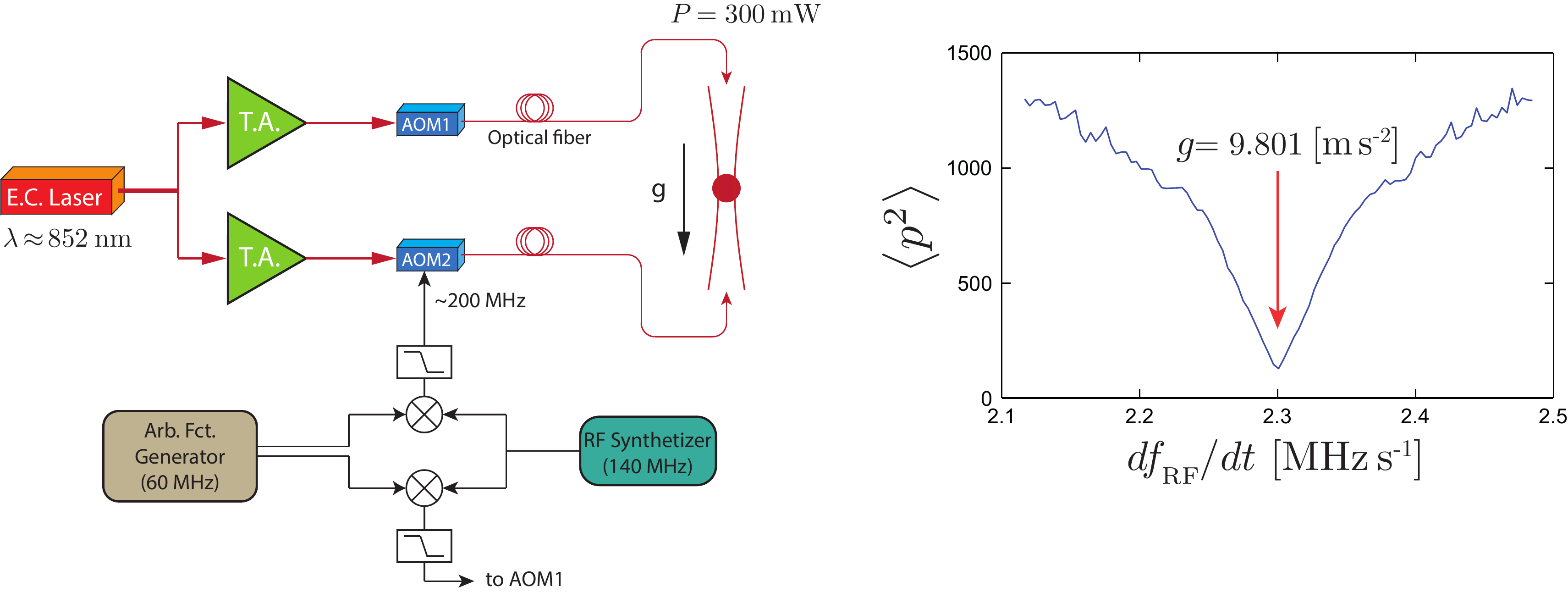}\caption{(\textit{left}) Schematic view of the vertical SW setup used for our Kicked Rotor experiment. The light of an External Cavity (EC) laser seeds two optical amplifiers and passes through two AOMs. The light is injected into two optical fibers and guided towards the interacting region. In addition to the pulsed sequence, a linear chirp is added to the driving RFs so that the SW ``falls'' simultaneously with the atomic cloud. (\textit{right}) Average square momentum $\left\langle p^{2}\right\rangle$ vs. the rate of the frequency chirp. The minimum corresponds to the optimal `compensation' of the gravity acceleration, where the SW nodes exactly follow the free-falling atomic cloud.}
	\label{fig:SW-scheme} 
\end{figure*}

\subsection{Experimental setup}
The atomic cloud
of cesium atoms is produced in a standard Magneto-Optical Trap (MOT), loaded during 1 s in a room-temperature atomic vapor cell. We then perform an optimized $55$ ms Sisyphus molasses phase which yields a few million atoms at a temperature of $\simeq2\,\mu$K. The MOT is switched off and the atomic cloud is ``kicked'' by a far-detuned ($\Delta\approx-10$ GHz at the cesium D2 line, with wavelength $\lambda_{L} = 852$~nm) pulsed standing wave, with a typical pulse duration of $\tau = 300$~ns. The kick frequency $1/T_{1}$ can be varied between $\sim$ 35 and 104 kHz, corresponding to $\kbar$ between 3 and 1. For our short values of $\tau$, the atomic motion is negligible during the application of the SW pulses, which can be considered as Dirac delta functions. After the kick sequence the momentum distribution of the cloud  is measured by a time-of-flight after a free fall of $\sim16$ cm from the MOT position.

\subsubsection{The standing wave}

The SW is formed by two independent counter propagating beams along
the vertical direction. The laser setup for the phase-modulated kicking
potential is similar to that of Ref.~\cite{White:ExperimentalQuantumRatchet:PRA13}, and is schematically shown in Fig.~\ref{fig:SW-scheme} (\textit{left}). A commercial external-cavity laser diode ($100$ kHz linewidth) beam is separated into two parts which seed two Masters Oscillator Power Amplifiers (MOPA) which yield Watt-range, mutually coherent beams. The amplified beams are sent through two acousto-optic modulators (AOMs) acting as fast switches (typical rise time of $15$ ns) which generate the pulses. An independent control of the phase and the amplitude of each beam is achieved using two separate, phase-locked radio frequency (RF) driving signals, which are modulated by an arbitrary wave function generator. The two resulting beams are injected in single-mode optical fibers ($8$~m long each) which transport the light to the interaction region. To reduce the SW phase noise we adjusted the optical path difference from the splitting point up to the atoms to 1 cm. The standing wave interacting with the atoms has a peak power of $300$ mW per beam and a waist $w_{0}=0.8$~mm.

Since the SW is parallel to gravity, the atoms are free-falling during
the application of the pulses. By applying a carefully-controlled
linear chirp ($df_{{\rm RF}}/dt={\rm const.}$, with $f_{\rm RF}$ the applied frequency) of opposite sign to each of the beam, one obtains a SW whose nodes follow the atoms during their free-fall. Thus, in the (non-inertial) reference frame of both the atoms and the SW we form a standard kicked rotor. Alternatively one can think that in the accelerated reference frame of the SW an inertial force exactly compensates gravity. This compensation is adjusted by minimizing the
measured average quadratic momentum $\langle p^{2}\rangle$ of the cloud at fixed number of kicks (typically 100 kicks, comparable to $t_\mathrm{loc}$ for our parameters) as a function of the relative frequency chirp. As a residual non-compensated acceleration breaks quasimomentum conservation and destroys dynamical localization, it  causes a sharp increase of $\langle p^{2}\rangle$, which displays a minimum around the optimal chirp value. This is shown in Fig.~\ref{fig:SW-scheme} (\textit{right}).

\subsubsection{Spatial filtering of the atomic cloud}

The QKR Hamiltonian Eq.~(\ref{eq:QKR}) has two parameters: the reduced
Planck constant $\kbar$ and the stochasticity parameter $K$. Both
are quite well controlled experimentally on relatively large ranges (1-3
for $\kbar$ , 0-17 for $K$). However,  a real cold-atom system is
much more complex, and depends on many other parameters. As
an example, consider the fact that whereas the model is 1D
the real system is 3D. As the momentum exchanges between atoms and the
SW are constrained along the SW direction the dynamics is effectively
1D (decoupled from the other directions). But the SW has a Gaussian transverse profile, which means that
the value of $K$ (depending on the beam intensity) varies along the
transverse extension of the beams. The spatial overlap between the
atomic density profile and the transverse intensity profile of the
SW beams is thus an additional parameter. A small ratio between
atomic cloud and SW beam sizes is desirable in order to reduce the $K$ spatial inhomogeneity. In our experiment, at the end of the Sisyphus molasses phase, the atomic cloud $e^{-2}$ radius $w_{\rm at}\sim 1$ mm and the SW beam waist is $w_{0}=0.8$ mm; this ratio is thus $\sim 1$. To avoid this problem, we implemented a spatial filtering of the atomic cloud in order to reduce its transverse extension (at the price of a large loss in the number of atoms), considerably reducing the inhomogeneity effects.

\begin{figure}[t!]
	\centering{}\centering \includegraphics[width=1\linewidth]{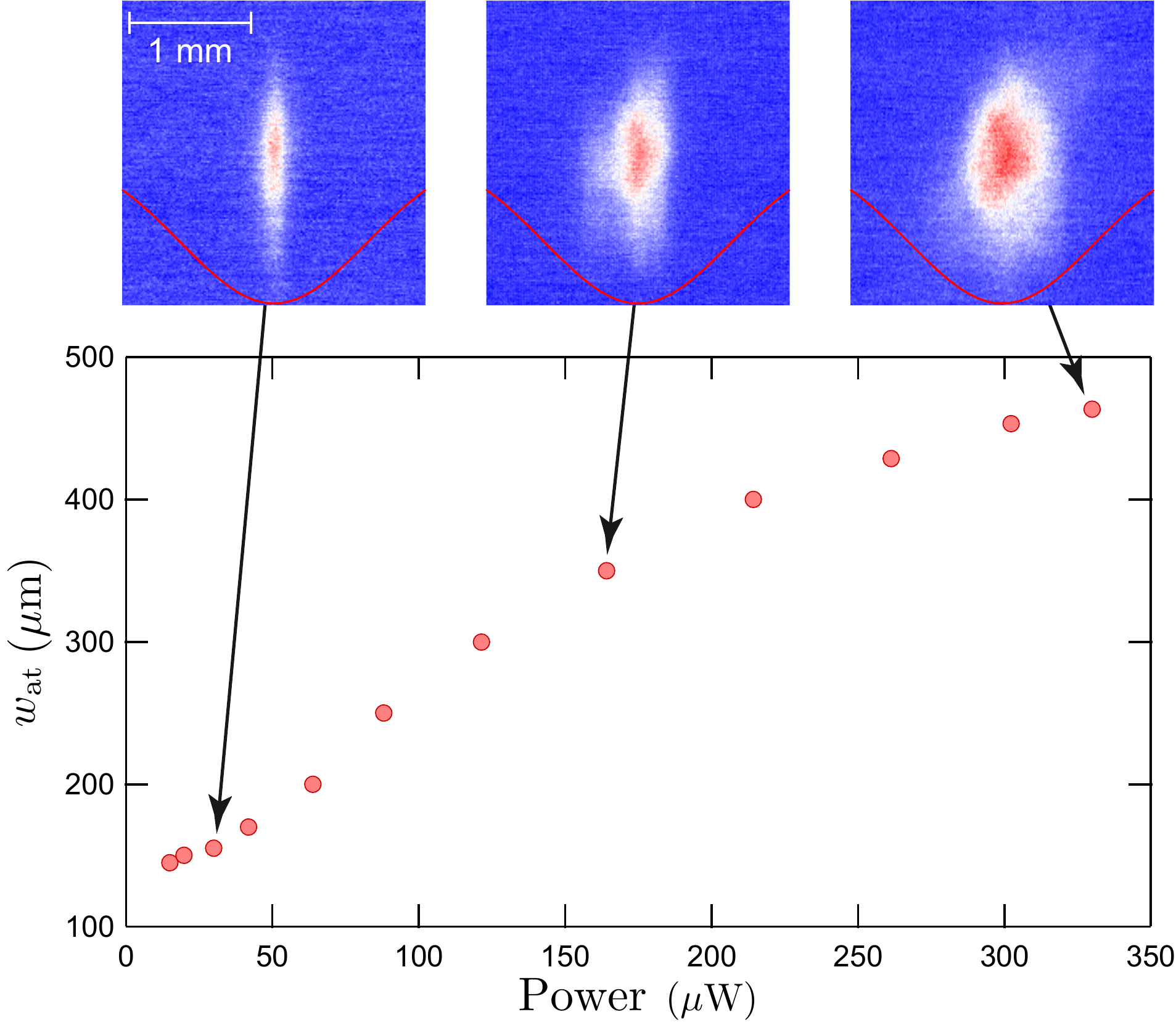}
	\caption{Dependence of the transverse atomic cloud size $w_{\mathrm{at}}$ on the ``small repumper'' beam power. The fluorescence images correspond to different filtered atomic clouds, with transverse sizes $w_{\mathrm{at}}$
of 150~$\mu$m, 340~$\mu$m and 470~$\mu$m respectively. The red curves are eye guides representing the Gaussian transverse profile of the SW beams  used in our experiment, with a waist of $w_{0}=800$~$\mu$m.}
	\label{fig:SmallRepumper} 
\end{figure}

Our filtering method uses a vertical, tightly-focused (100~$\mu$m waist) ``repumper'' beam, carefully aligned with the axis of the SW. At the very beginning of the molasses (working on the $F_{g}=4\rightarrow F_{e}=5$ hyperfine transition) phase,
the MOT standard repumper beam, tuned to the $F_{g}=3\rightarrow F_{e}=4$ transition is switched off, and the ``small'', focused repumper (on the same transition) is switched on. The atoms outside the volume delimited by this beam eventually escape the optical cycling by falling into the $F_{g}=3$ level and are not affected anymore by the molasses beams. In order to remove these undesirable atoms, a horizontal pusher beam resonant with the $F_{g}=3\,\rightarrow F_{e}=2$ transition is applied to the entire cloud but does not affect the ``useful''
$F_{g}=4$ atoms, with which it is not resonant. We found empirically that the best filtering efficiency is obtained with 50~$\mu$W power and a 100 MHz detuning. The strong atom loss ($\sim 90\%$) induced by the filtering process can be mitigated by increasing the averaging over the number of experimental realizations. Additional advantages of the filtering are a reduction of the fluctuations in size and the position of the cloud, as well as a precise control of its size. Figure~\ref{fig:SmallRepumper} shows the dependence of the transverse size of the cloud $w_{\rm at}$ vs. the filtering beam power, as well as a few fluorescence images corresponding to different cloud sizes.

\section{Diffusion coefficient for the PSQKR \label{sec:Dini}}

The above-described experimental setup was used to investigate the short-time transport properties of our system. For that, we performed a quantitative study of the PSQKR compared to the standard QKR and the RKR, and we showed that the effects of the correlations between kicks can be suppressed by averaging over the PSQKR phase shifts ${a_{j}}$.

\subsection{Effect of correlations on the diffusion coefficient}

The diffusion coefficient is defined as $D_{\textrm{ini}}=[\langle p^{2}(t)\rangle/2t]_{t\ll t_{\mathrm{loc}}}$. For perfectly uncorrelated kicks, as in the RKR case, the value of the diffusion coefficient is $D_{0}=K^{2}/4$~\cite{Rechester:KRDiffCoeff:PRA81} (where, for convenience, we express the momentum in units of the width of the Brillouin zone,  $2\hbar k_{L}$, in the units of Eq.~(\ref{eq:QKR}) it reads $K^2/4\kbar$). In presence of correlations, it is well-known that oscillations appear as the kick strength $K$ is varied, both in
the classical and quantum cases.

\begin{figure}[t!]
	\centering{}\includegraphics[width=1\linewidth]{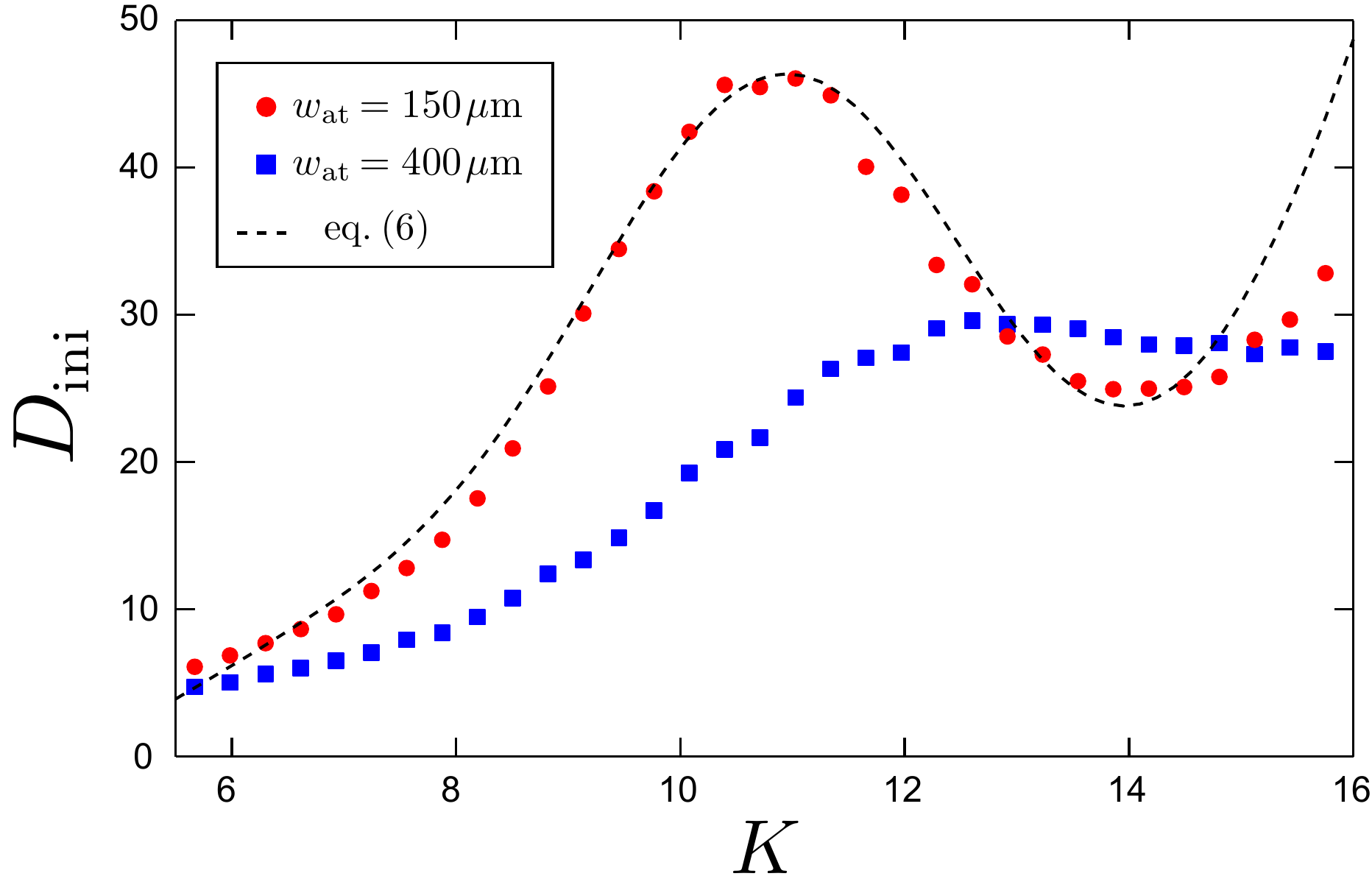}\caption{Measurement of the initial diffusion coefficient $D_{\mathrm{ini}}$ vs. the kick strength $K$ for the standard QKR. The dashed black line is Eq.~\eqref{eq:DiniQKR}  and the symbols represent measurements for two different values of $w_{\mathrm{at}}$. For $w_{\mathrm{at}}=150$~$\mu\mathrm{m}\ll w_{0}$ (red circles) we find a good agreement with the theoretical curve, whereas for $w_{\mathrm{at}}=400$~$\mu$m (blue rectangles) we observe a significant deviation due to inhomogeneity of $K$. All data was taken using $\kbar=2.89$, and $D_{\mathrm{ini}}$ was obtained by measuring $\left\langle p^{2}\right\rangle $ after $t_0=7$ kicks.}
	\label{fig:Dini_vs_K} 
\end{figure}

The oscillations of $D_{\textrm{ini}}$ in the classical and quantum kicked rotor
have been previously studied both theoretically \cite{Rechester:KRDiffCoeff:PRA81,Rechester:TurbulentDiffChirikovTaylor:PRL80,Shepelyansky:Bicolor:PD87}
and experimentally \cite{Raizen:KRClassRes:PRL98}. In particular
the dependence of the diffusion coefficient with $K$ for the standard QKR has
been calculated in~\cite{Shepelyansky:Bicolor:PD87} and is given
by 
\begin{equation}
\begin{split}D_{\text{ini}}=D_{0}( & 1-2J_{2}(K_{q})-2J_{1}^{2}(K_{q})\\
 & +2J_{3}^{2}(K_{q})+2J_{2}^{2}(K_{q})\ldots),
\end{split}
\label{eq:DiniQKR}
\end{equation}
where $J_{n}(z)$ is the Bessel function of the first kind of order $n$, and $K_{q}=(2K/\kbar)\sin(\kbar/2)$
is an effective kick strength taking into account quantum effects, i.e. the non-commutative nature
of the position and momentum operators. The main correction is $-2J_{2}(K_{q})$,
and comes from the correlations of the momentum between kicks $k$ and $k+2$. The next corrections
$-2J_{1}^{2}(K_{q})+2J_{3}^{2}(K_{q})$ and $2J_{2}^{2}(K_{q})$ come from correlations
at three and four kicks respectively. The effects of higher-order correlations (as well as subleading corrections to four-kick correlations) can be neglected for our present purposes.

\begin{figure}[t!]
	\centering{}\includegraphics[width=1\linewidth]{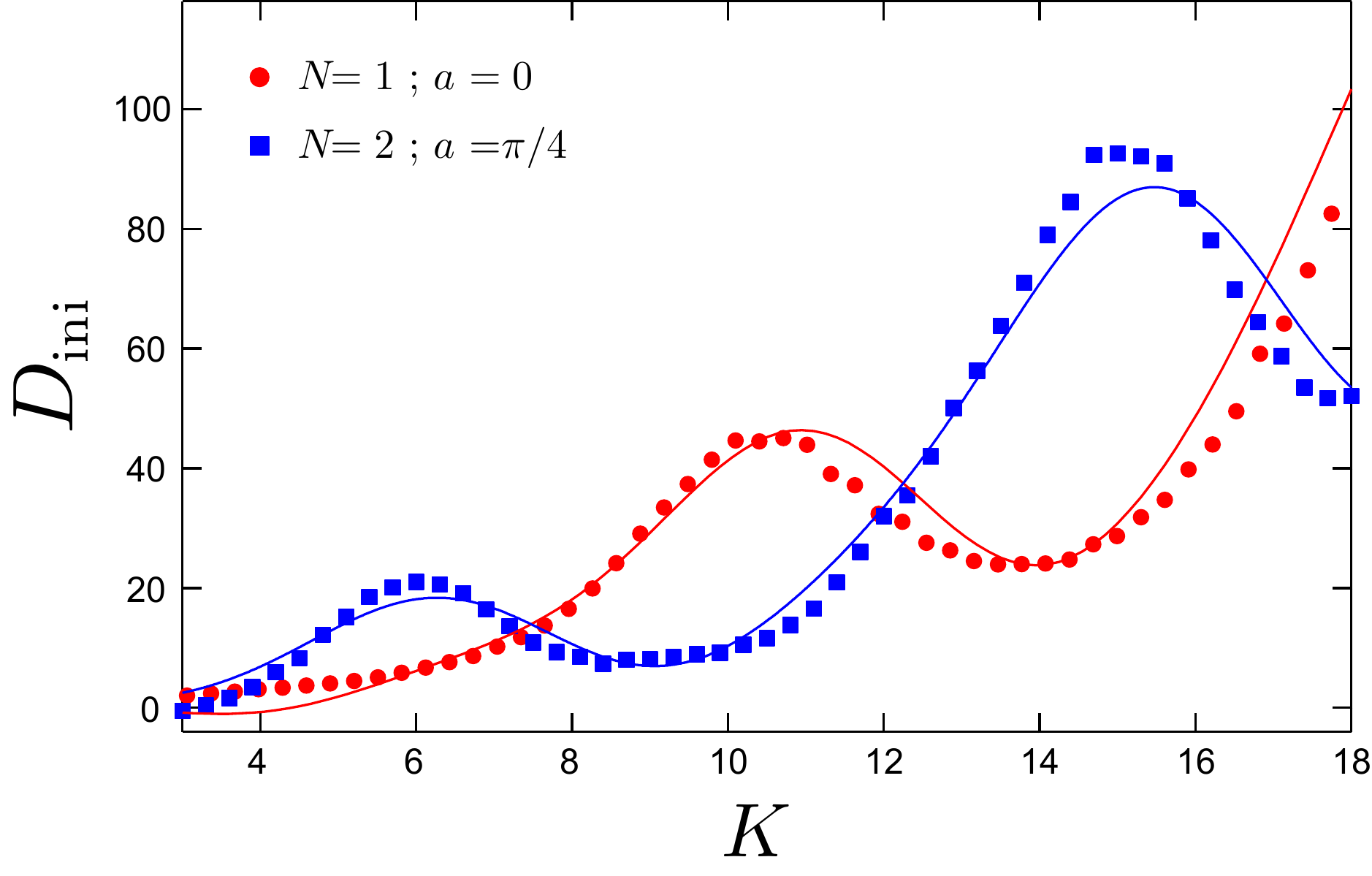}
	\caption{Measured diffusion coefficient for the standard QKR [$N=1,\,a=0$, red circles, the full line corresponds to Eq.~(\ref{eq:DiniQKR})] compared to the period-two PSQKR [$N=2,\,a=\pi/4$, blue squares, the full line corresponds to Eq.~(\ref{eq:DiniPSQKR})]. We observe that the oscillations (essentially due to first order correlation terms) are shifted almost to be opposite to each other. $D_{\text{ini}}$ was determined by measuring $\left\langle p^{2}\right\rangle $ after $t_0=7$ kicks and $\kbar=2.89$.} 
	\label{fig:Dini_vs_a} 
\end{figure}

The derivation of Ref.~\cite{Rechester:KRDiffCoeff:PRA81,Shepelyansky:Bicolor:PD87} can be generalized
to the diffusion coefficient $D_{\mathrm{ini}}$ of the PSQKR. In the simplest cases
$N=2$ and $N=3$ we obtain the diffusion
coefficient $D_{\text{ini},N}$: 
\begin{equation}
\begin{split}D_{\text{ini},2}=D_{0} \{ & 1-2J_{2}(K_{q})\cos(2(a_{2}-a_{1}))-2J_{1}^{2}(K_{q})\\
 & +2J_{3}^{2}(K_{q})\cos(4(a_{2}-a_{1}))+2J_{2}^{2}(K_{q}) \ldots \}\\
D_{\text{ini},3}=D_{0} \{ & 1-2C_{1}[J_{2}(K_{q})+J_{1}^{2}(K_{q})]\\
 & +2C_{2}[J_{3}^{2}(K_{q})+J_{2}^{2}(K_{q})]+\ldots \},
\end{split}
\label{eq:DiniPSQKR}
\end{equation}
where 
\begin{eqnarray}
C_{1}&=&\frac{1}{3}\Big(\cos(a_{1}-2a_{2}+a_{3})+\cos(a_{2}-2a_{3}+a_{1})\\
&\quad \quad &\quad \quad +\cos(a_{3}-2a_{1}+a_{2})\Big) \\
C_{2}&=&\frac{1}{3}\left(\cos(3a_{2}-3a_{1})+\cos(3a_{3}-3a_{2})+\cos(3a_{1}-3a_{3})\right)\,.
\end{eqnarray}
Here also, the term proportional to $J_{2}(K_{q})$ comes from the two-kick correlation, while the terms proportional to $-2J_{1}^{2}(K_{q})$ and $2J_{3}^{2}(K_{q})$ ($2J_{2}^{2}(K_{q})$) come from correlations
at three (four) kicks.
The corrections to $D_{\mathrm{ini}}$ depend on the phases $a_{j}$, and the choice of the phase-modulation sequence will strongly affect kick correlations in the PSQKR. A carefully tailored sequence $a_j$ might thus provide a method to control the dependence of diffusion coefficient on $K_{q}$.

\begin{figure}[t!]
	\centering{}\includegraphics[width=1\linewidth]{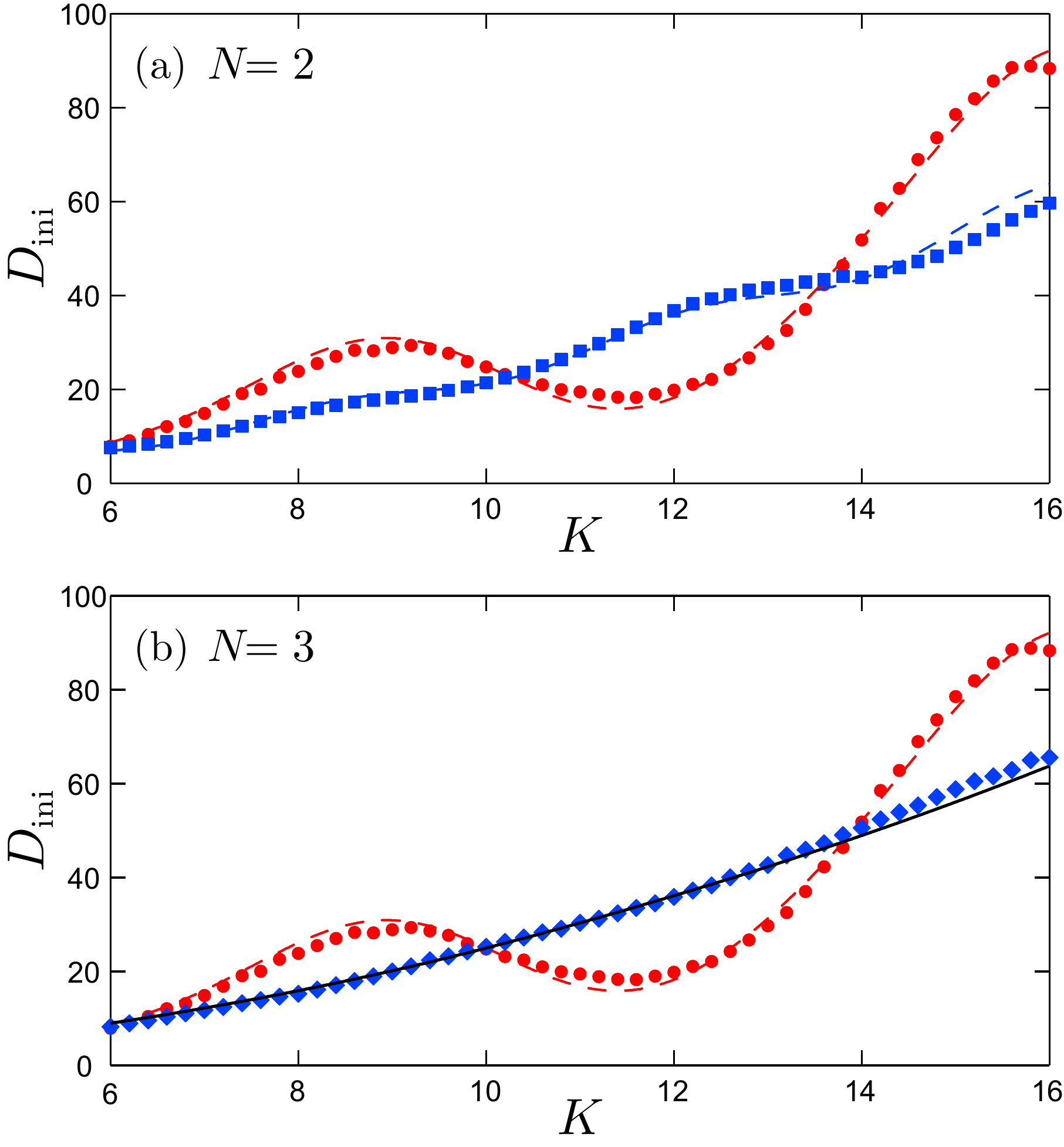}
	\caption{Comparison between the mesured diffusion coefficient for the QKR (red circles) and
		the phase-averaged PSQKR [(a) $N=2$, blue squares; (b) $N=3$ blue diamonds] with $\kbar=2$ and $D_{{\rm ini}}$ extracted after 7 kicks. Plot (a) shows a moderate smearing of the correlation-induced oscillations due to the phase averaging with the period 2 PSQKR, plot (b) shows the much more complete smearing obtained with period 3. The dashed lines are the predictions of Eq.~\eqref{eq:DiniPSQKR}, and the solid line in plot (b) is the RKR (no correlation) prediction $D_0=K^2/4$.}
	\label{fig:CorrelSmoothing} 
\end{figure}

In order to test this prediction, we experimentally measured the initial diffusion coefficient $D_{\textrm{ini}}$. This measurement is done by analyzing the time-of-flight atomic distribution after a small number of kicks (typically $\sim 7$). In these conditions, we observe that the distributions preserve a Gaussian shape (confirming that the dynamics is still diffusive), and that the square of the fitted width of the momentum distribution, $\left\langle p^{2}(t_0)\right\rangle\equiv\sigma_{p}^{2}(t)-\sigma_{p}^{2}(0)$, varies linearly with the kick number $t$. From the measured width at a given number of kicks $t_0 \sim 7$  we extract $\sigma_{p}^2(t_0)$, and thus infer the diffusion coefficient $D_{\textrm{ini}}=\left\langle p^{2}(t_0)\right\rangle/2t_0$.

\begin{figure*}[t!]
	\centering{}\includegraphics[width=1\linewidth]{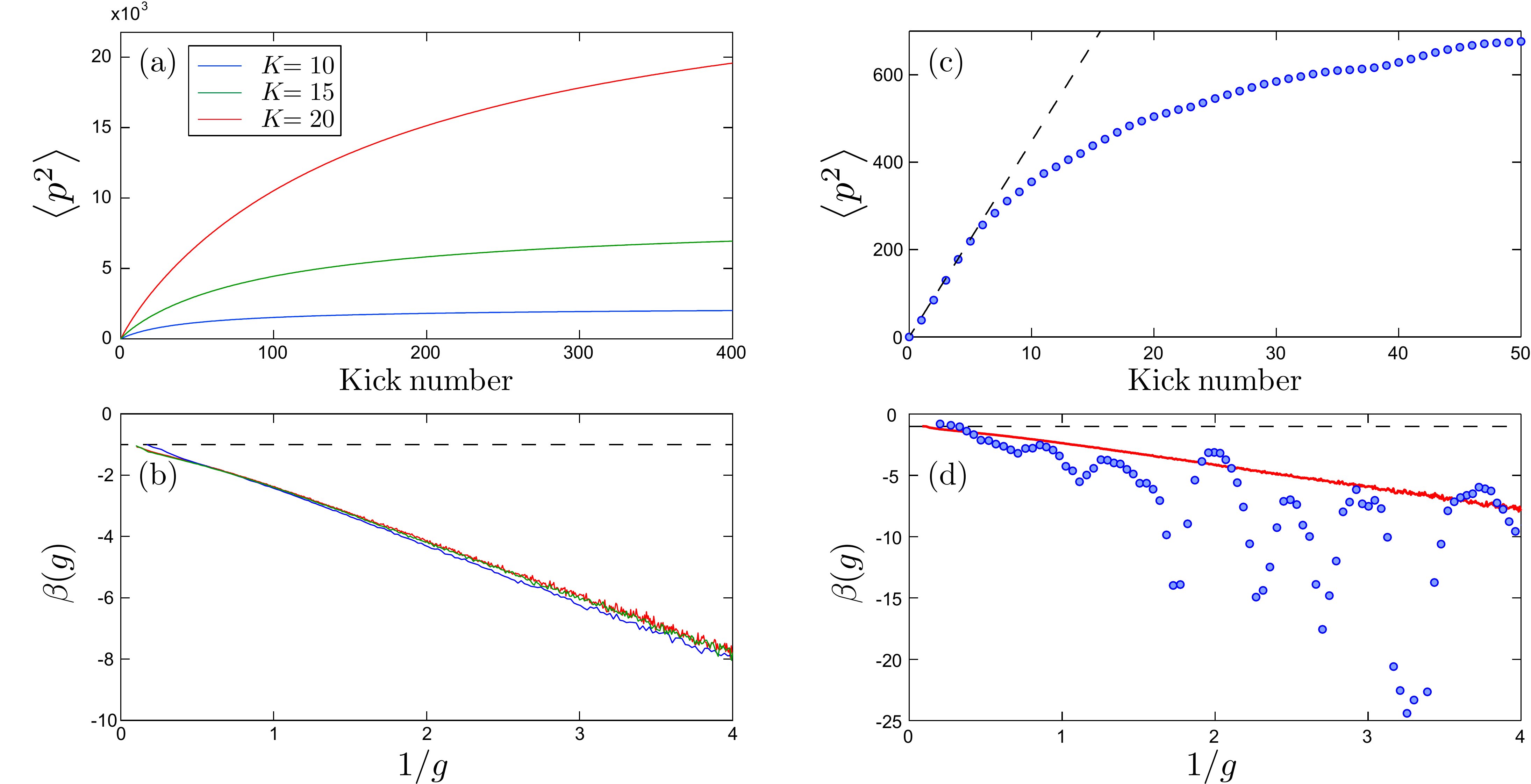}
	\caption{Numerical simulation of $\langle p^{2}(t)\rangle$ and the corresponding $\beta(g)$ functions for the RKR and QKR. (a) RKR's $\langle p^{2}(t)\rangle$ for three values of the parameter $K= 10,15,20$ (equivalent to  $K/\kbar$ for the QKR), the curves are very smooth and correlation free. (b) The corresponding $\beta(g)$ functions coincide, evidencing the universal character of $\beta(g)$. (c) Standard QKR's $\langle p^{2}(t)\rangle$ (blue circles) and initial diffusion (black dashed line),  with $K=5.8$ and $\kbar=0.8$. The correlation-induced oscillations, although small, are clearly visible. (d) The corresponding $\beta(g)$ (blue circles) showing the amplification of these oscillations due to the logarithmic derivative. These oscillations lead to large deviations with respect to the universal scaling function, materialized by the RKR result (solid red line).}
	\label{fig:BetaSimpleKR} 
\end{figure*}

Figure~\ref{fig:Dini_vs_K} displays the measured dependence of $D_{\textrm{ini}}$ vs. $K$ for the standard QKR and compares it with Eq.~\eqref{eq:DiniQKR}, in the presence and in the absence of spatial filtering of the atomic cloud described in Sec.~\ref{sec:PSQKR}. The effect of filtering  in reducing the transverse inhomogeneity appears to be very  important. For an atomic cloud size $w_{\rm at}=400\,\mu$m (blue rectangles), about the half of the SW waist ($w_0 =800\,\mu$m), the oscillations are severely smeared. When the filtering procedure is used to reduce the atom cloud size to $w_{\textrm{at}}=150$ $\mu$m, we clearly observe the oscillations induced by kick correlations (red circles), in good agreement with Eq.~\eqref{eq:DiniQKR} (dashed black line). For $w_{\textrm{at}}<200\,\mu$m) the curve becomes almost independent of $w_{\textrm{at}}$, indicating that residual inhomogeneity is negligible.

Figure~\ref{fig:Dini_vs_a} shows the same measurement of $D_\text{ini}$ vs. $K$ for the PSQKR in the simplest case, $N=2$, with the phase-shift sequence $a_{1}=a$, $a_{2}=-a$, with   $a=\pi/4$ (blue rectangles). The solid lines correspond to Eq.~\eqref{eq:DiniPSQKR}, and are in very good agreement with the experimental data. The main oscillations are seen to be opposite to the ones of the standard QKR (red circles). This is due to the fact that the leading term in Eq.~\eqref{eq:DiniPSQKR} has the same amplitude as for the standard QKR, but its sign is inverted.  

\subsection{Suppression of the correlation effects by the averaging over the phases}

The above results suggest that one can use the effect of the PSQKR phase shifts on the kick correlations provided to suppress these undesirable effects. 
We will show that it is possible to average over randomly-chosen phase sequences in order
to mimic the behavior of an ideal (correlation-free) disordered system,
corresponding to the diffusion coefficient of the correlationless
RKR, $D_{0}=K^{2}/4$. This can be achieved even for relatively low modulation periods, as the effect of higher-order correlations remains small, as one can see from Eq.~\eqref{eq:DiniPSQKR}. With $N=2$, one can eliminate the two-kick correlations ($\propto J_2(K_q)$), and part of the three-kick correlations  ($\propto J_3^2(K_q)$). 
With $N=3$, one can eliminate all terms in $J_{n}^{2}$.

The experimental results are shown in Fig.~\ref{fig:CorrelSmoothing} for $N=2$, plot (a), and $N=3$, plot
(b). In both panels, the red circles and line correspond resp.
to experimental data and Eq.~\eqref{eq:DiniQKR} for the
standard QKR (with $\kbar=2$), where all the correlation terms are
present. Blue symbols correspond to the PSQKR with the experimental
results averaged over 100 different sets of phases $a_{j}$, while the lines correspond to the theoretical results, Eq.~\eqref{eq:DiniPSQKR}, averaged over the phases.
The agreement between theory and experiment is very good, and one
clearly observes a reduction of the oscillations due to the averaging.
In the case $N=3$ [plot (b)], the experimental data very well agrees with the uncorrelated RKR result $D_{0}=K^{2}/4$ (solid line), evidencing the efficiency of the averaging for suppressing the effect of correlations.

\begin{figure}[t!]
	\centering{} \includegraphics[width=1\linewidth]{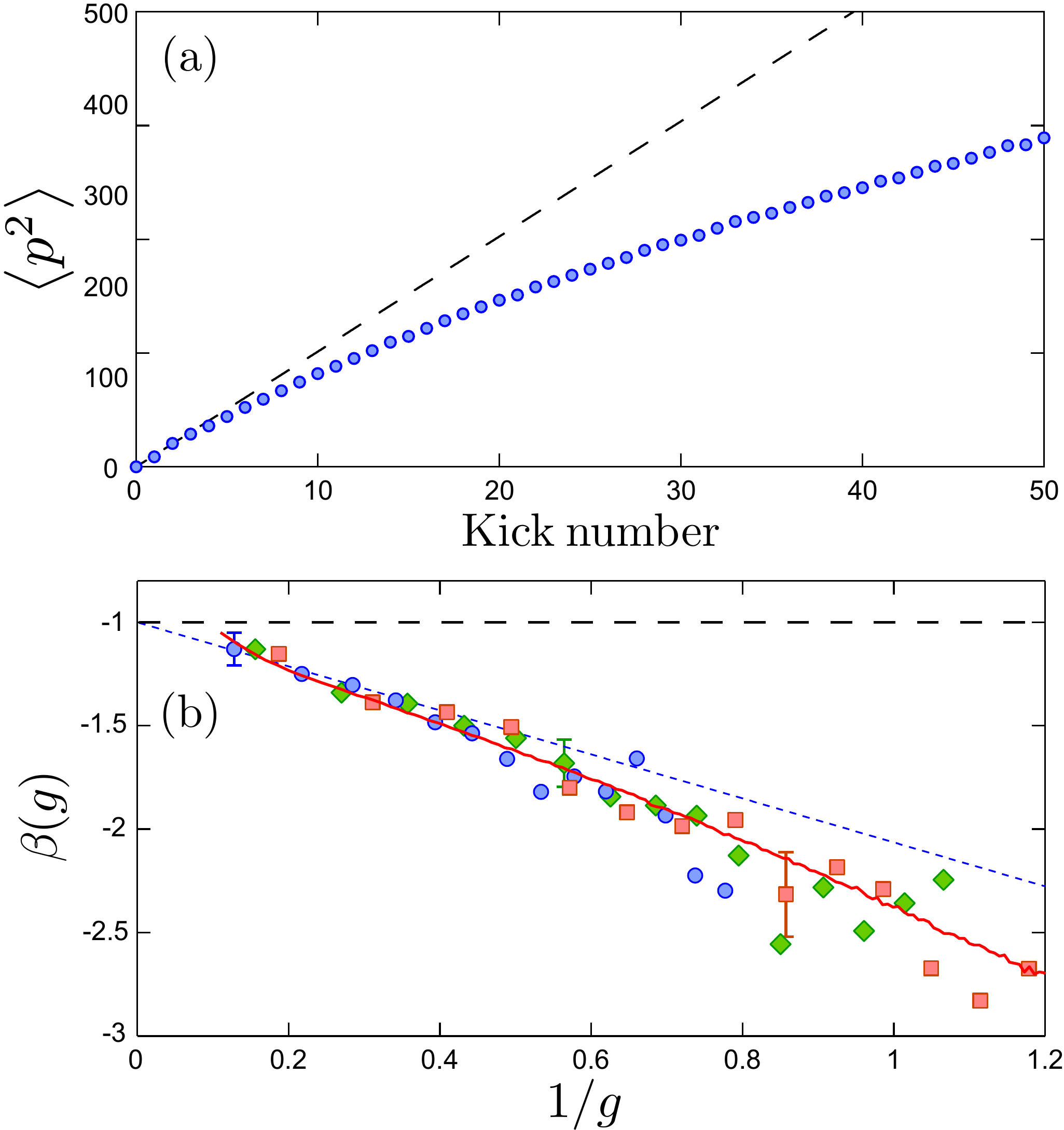}
	\caption{(a)
		Experimentally measured $\langle p^{2}(t)\rangle$ for the PSQKR ($N=3$, $K=4$). The dashed line is a linear fit at low kick number (0 to 5 kicks), corresponding to the initial diffusive behavior. (b) The experimental one-parameter scaling function $\beta(g)$, for three different parameter values: ($N=3$, $K=4$) -- blue circles, ($N=4$, $K=4.5$) -- green diamonds and ($N=5$, $K=3.5$) -- red squares. These results are in excellent agreement with RKR numerical simulations (red solid line) and, in the large-conductance regime $1/g\le0.3$, in good agreement with the diagrammatic theory prediction for the orthogonal class: $\beta(g)\simeq -1-1.064/g$ (dashed blue line)~\cite{Tian:EhrenfestTimeDynamicalLoc:PRB05}. $\kbar=1$.}
	\label{fig:BetaPSKR_expe} 
\end{figure}

\section{Measurement of the scaling function $\beta(g)$}
\label{sec_beta}

In this section we show how the averaging over phase-shifts allows for a clean measurement of the universal one-parameter scaling function $\beta(g)$. We consider only the orthogonal symmetry class, to which the standard QKR belongs \footnote{The realization of a PSQKR in the unitary symmetry class is demonstrated in Ref.~\cite{Hainaut:CFS:NCM18}}. In order to realize this case with the PSQKR, we have to constrain the phase-shifts sequence by an anti-symmetry condition~\cite{Hainaut:CFS:NCM18}.

In Figure~\ref{fig:BetaSimpleKR}(a) we show  a numerical simulation of $\langle p^{2}(t)\rangle$ for the RKR, for three different values of the kick amplitude, $K=10, 15$ and 20. The curves are smooth and do not show, as expected, any correlation effects. The universal character of the dynamics (i.e. its independence of the ``microscopic'' parameter $K$) is best evidenced by the collapse onto one curve of the $\beta(g)$ function [computed using Eq.~\eqref{eq:betaQKR}], as shown in Fig.~\ref{fig:BetaSimpleKR}(b). Fig.~\ref{fig:BetaSimpleKR}(c) shows  a numerical simulation of $\langle p^{2}(t)\rangle$ for the standard QKR (blue circles),  using  typical parameter values accessible in experiments ($K=5.8$ and $\kbar=0.8$). The dashed black line is an extrapolation of the initial diffusion, followed by the beginning of dynamical localization, common with the RKR behavior, but we clearly observe small oscillations due to temporal correlations. Figure~\ref{fig:BetaSimpleKR}(d) shows that the corresponding $\beta(g)$ function is strongly affected by these non-universal oscillations, masking its universal character (indicated by the red solid line corresponding to the RKR behavior with the same parameters).

Figure~\ref{fig:BetaPSKR_expe} shows the experimentally measured $\langle p^{2}(t)\rangle$ for the PSQKR [plot (a)] and the corresponding $\beta(g)$ function [plot (b)]. Three sets of experiments were used, with different parameters values: $N=3$, $K=4$ (blue circles);  $N=4$, $K=4.5$ (green diamonds); $N=5$, $K=3.5$ (red squares ); all with $\kbar=1$. The experimental momentum distributions were averaged over 100 randomly-generated sets of phase shifts (constrained by the anti-symmetry condition). A careful analysis was used for extracting $\langle p^{2}(t)\rangle$ from the time-of-flight momentum distributions $\Pi(p)$. Since $\langle p^{2}(t)\rangle$ is sensitive to the tails of the momentum distribution, we fit $p^2 \Pi(p)$ rather than $\Pi(p)$ itself. The fitting function is the Lobkis-Weaver self-consistent function~\cite{LobkisWeaver:SelfConsistentTransportLocWaves:PRE05}, which smoothly evolves from a Gaussian to an exponential shape. This method was tested and validated with numerical simulations over a wide range of parameters. Plot (a) shows that although relatively low kick strengths were used, similar to the ones in Fig.~\ref{fig:BetaSimpleKR}(c), $\langle p^{2}(t)\rangle$ is very smooth, with no visible oscillations (to improve the figure clarity we displayed only the curve corresponding to the first set of parameters). Fig.~\ref{fig:BetaPSKR_expe}(b) shows the corresponding $\beta(g)$ functions, in good  agreement with the Random Kicked Rotor scaling (solid red line). The agreement among all these curves (to within experimental errors) is a clear experimental proof of the one-parameter scaling function universality, and evidences the efficiency of the phase-averaging method to suppress temporal correlations.

\section{Conclusion \label{sec_concl}}

In conclusion, we described in this work powerful techniques, both experimental and for data analysis, allowing us to obtain high quality measurements of important parameters, like the initial diffusion coefficient. The most important of these techniques is the implementation of an engineered Hamiltonian, the phase-shifted quantum kicked rotor. We showed how these improved setup and analysis can suppress the effects of kick correlations and allowed
us to perform the quite challenging task of measuring a central quantity
characterizing quantum transport in disordered systems, the celebrated
$\beta(g)$ function. This considerably enhances the potential of
kicked cold-atomic systems for quantum simulations of disordered
systems, and in particular their transport properties, as put in evidence
in our recent works~\cite{Hainaut:CFS:NCM18,Hainaut:EnhancedReturnOrigin:PRL17,Hainaut:RatchetEffectQKR:PRA18,Hainaut:TimePhaseTrQuChaos:PRL18}.

\subsection*{ACKNOWLEDGMENTS}

This work was supported by Agence Nationale de la Recherche (Grants
LAKRIDI No. ANR-11-BS04-0003 and K-BEC No. ANR-13-BS04-0001-01), the
Labex CEMPI (Grant No. ANR-11-LABX-0007-01), and ``Fonds Europ{\'e}en
de D{\'e}veloppement {\'E}conomique R{\'e}gional'' through the ``Programme Investissements
d'Avenir''.


\begin{thebibliography}{44}%
\makeatletter
\providecommand \@ifxundefined [1]{%
 \@ifx{#1\undefined}
}%
\providecommand \@ifnum [1]{%
 \ifnum #1\expandafter \@firstoftwo
 \else \expandafter \@secondoftwo
 \fi
}%
\providecommand \@ifx [1]{%
 \ifx #1\expandafter \@firstoftwo
 \else \expandafter \@secondoftwo
 \fi
}%
\providecommand \natexlab [1]{#1}%
\providecommand \enquote  [1]{``#1''}%
\providecommand \bibnamefont  [1]{#1}%
\providecommand \bibfnamefont [1]{#1}%
\providecommand \citenamefont [1]{#1}%
\providecommand \href@noop [0]{\@secondoftwo}%
\providecommand \href [0]{\begingroup \@sanitize@url \@href}%
\providecommand \@href[1]{\@@startlink{#1}\@@href}%
\providecommand \@@href[1]{\endgroup#1\@@endlink}%
\providecommand \@sanitize@url [0]{\catcode `\\12\catcode `\$12\catcode
  `\&12\catcode `\#12\catcode `\^12\catcode `\_12\catcode `\%12\relax}%
\providecommand \@@startlink[1]{}%
\providecommand \@@endlink[0]{}%
\providecommand \url  [0]{\begingroup\@sanitize@url \@url }%
\providecommand \@url [1]{\endgroup\@href {#1}{\urlprefix }}%
\providecommand \urlprefix  [0]{URL }%
\providecommand \Eprint [0]{\href }%
\providecommand \doibase [0]{http://dx.doi.org/}%
\providecommand \selectlanguage [0]{\@gobble}%
\providecommand \bibinfo  [0]{\@secondoftwo}%
\providecommand \bibfield  [0]{\@secondoftwo}%
\providecommand \translation [1]{[#1]}%
\providecommand \BibitemOpen [0]{}%
\providecommand \bibitemStop [0]{}%
\providecommand \bibitemNoStop [0]{.\EOS\space}%
\providecommand \EOS [0]{\spacefactor3000\relax}%
\providecommand \BibitemShut  [1]{\csname bibitem#1\endcsname}%
\let\auto@bib@innerbib\@empty
\bibitem [{\citenamefont {Anderson}(1958)}]{Anderson:LocAnderson:PR58}%
  \BibitemOpen
  \bibfield  {author} {\bibinfo {author} {\bibfnamefont {P.~W.}\ \bibnamefont
  {Anderson}},\ }\bibfield  {title} {\enquote {\bibinfo {title} {{Absence of
  Diffusion in Certain Random Lattices}},}\ }\href {\doibase
  10.1103/PhysRev.109.1492} {\bibfield  {journal} {\bibinfo  {journal} {Phys.
  Rev.}\ }\textbf {\bibinfo {volume} {109}},\ \bibinfo {pages} {1492--1505}
  (\bibinfo {year} {1958})}\BibitemShut {NoStop}%
\bibitem [{\citenamefont {Mazza}\ \emph {et~al.}(2012)\citenamefont {Mazza},
  \citenamefont {Bermudez}, \citenamefont {Goldman}, \citenamefont {Rizzi},
  \citenamefont {Martin-Delgado},\ and\ \citenamefont
  {Lewenstein}}]{Mazza:OpticalLatticeBasedQuantumSimulator:NJP12}%
  \BibitemOpen
  \bibfield  {author} {\bibinfo {author} {\bibfnamefont {L.}~\bibnamefont
  {Mazza}}, \bibinfo {author} {\bibfnamefont {A.}~\bibnamefont {Bermudez}},
  \bibinfo {author} {\bibfnamefont {N.}~\bibnamefont {Goldman}}, \bibinfo
  {author} {\bibfnamefont {M.}~\bibnamefont {Rizzi}}, \bibinfo {author}
  {\bibfnamefont {M.~A.}\ \bibnamefont {Martin-Delgado}}, \ and\ \bibinfo
  {author} {\bibfnamefont {M.}~\bibnamefont {Lewenstein}},\ }\bibfield  {title}
  {\enquote {\bibinfo {title} {{An optical-lattice-based quantum simulator for
  relativistic field theories and topological insulators}},}\ }\href
  {{stacks.iop.org/1367-2630/14/i=1/a=015007}} {\bibfield  {journal} {\bibinfo
  {journal} {New J. Phys}\ }\textbf {\bibinfo {volume} {14}},\ \bibinfo {pages}
  {015007} (\bibinfo {year} {2012})}\BibitemShut {NoStop}%
\bibitem [{\citenamefont {Bloch}\ \emph {et~al.}(2014)\citenamefont {Bloch},
  \citenamefont {Dalibard},\ and\ \citenamefont
  {Nascimbene}}]{Bloch:QuantumSimulationsUltracoldGases:NP14}%
  \BibitemOpen
  \bibfield  {author} {\bibinfo {author} {\bibfnamefont {I.}~\bibnamefont
  {Bloch}}, \bibinfo {author} {\bibfnamefont {J.}~\bibnamefont {Dalibard}}, \
  and\ \bibinfo {author} {\bibfnamefont {S.}~\bibnamefont {Nascimbene}},\
  }\bibfield  {title} {\enquote {\bibinfo {title} {{Quantum simulations with
  ultracold quantum gases}},}\ }\href {\doibase 10.1038/nphys2259} {\bibfield
  {journal} {\bibinfo  {journal} {Nat. Phys.}\ }\textbf {\bibinfo {volume}
  {8}},\ \bibinfo {pages} {267--276} (\bibinfo {year} {2014})}\BibitemShut
  {NoStop}%
\bibitem [{\citenamefont {Georgescu}\ \emph {et~al.}(2014)\citenamefont
  {Georgescu}, \citenamefont {Ashhab},\ and\ \citenamefont
  {Nori}}]{Georgescu:QuantumSimulation:RMP14}%
  \BibitemOpen
  \bibfield  {author} {\bibinfo {author} {\bibfnamefont {I.~M.}\ \bibnamefont
  {Georgescu}}, \bibinfo {author} {\bibfnamefont {S.}~\bibnamefont {Ashhab}}, \
  and\ \bibinfo {author} {\bibfnamefont {F.}~\bibnamefont {Nori}},\ }\bibfield
  {title} {\enquote {\bibinfo {title} {{Quantum simulation}},}\ }\href
  {\doibase 10.1103/RevModPhys.86.153} {\bibfield  {journal} {\bibinfo
  {journal} {Rev. Mod. Phys.}\ }\textbf {\bibinfo {volume} {86}},\ \bibinfo
  {pages} {153--185} (\bibinfo {year} {2014})}\BibitemShut {NoStop}%
\bibitem [{\citenamefont {Feynman}(1982)}]{Feynman:SimulatingPhysics:IJTP82}%
  \BibitemOpen
  \bibfield  {author} {\bibinfo {author} {\bibfnamefont {R.~P.}\ \bibnamefont
  {Feynman}},\ }\bibfield  {title} {\enquote {\bibinfo {title} {{Simulating
  Physics with Computers}},}\ }\href@noop {} {\bibfield  {journal} {\bibinfo
  {journal} {Int. J. Theor. Phys.}\ }\textbf {\bibinfo {volume} {21}},\
  \bibinfo {pages} {467--488} (\bibinfo {year} {1982})}\BibitemShut {NoStop}%
\bibitem [{\citenamefont {Bloch}\ \emph {et~al.}(2008)\citenamefont {Bloch},
  \citenamefont {Dalibard},\ and\ \citenamefont
  {Zwerger}}]{Bloch:ManyBodyUltracold:RMP08}%
  \BibitemOpen
  \bibfield  {author} {\bibinfo {author} {\bibfnamefont {I.}~\bibnamefont
  {Bloch}}, \bibinfo {author} {\bibfnamefont {J.}~\bibnamefont {Dalibard}}, \
  and\ \bibinfo {author} {\bibfnamefont {W.}~\bibnamefont {Zwerger}},\
  }\bibfield  {title} {\enquote {\bibinfo {title} {{Many-body physics with
  ultracold gases}},}\ }\href {\doibase 10.1103/RevModPhys.80.885} {\bibfield
  {journal} {\bibinfo  {journal} {Rev. Mod. Phys.}\ }\textbf {\bibinfo {volume}
  {80}},\ \bibinfo {pages} {885--964} (\bibinfo {year} {2008})}\BibitemShut
  {NoStop}%
\bibitem [{\citenamefont {Greiner}\ \emph {et~al.}(2002)\citenamefont
  {Greiner}, \citenamefont {Mandel}, \citenamefont {Esslinger}, \citenamefont
  {H{\"a}nsch},\ and\ \citenamefont {Bloch}}]{Greiner:MottTransition:N02}%
  \BibitemOpen
  \bibfield  {author} {\bibinfo {author} {\bibfnamefont {M.}~\bibnamefont
  {Greiner}}, \bibinfo {author} {\bibfnamefont {O.}~\bibnamefont {Mandel}},
  \bibinfo {author} {\bibfnamefont {T.}~\bibnamefont {Esslinger}}, \bibinfo
  {author} {\bibfnamefont {T.~W.}\ \bibnamefont {H{\"a}nsch}}, \ and\ \bibinfo
  {author} {\bibfnamefont {I.}~\bibnamefont {Bloch}},\ }\bibfield  {title}
  {\enquote {\bibinfo {title} {{Quantum phase transition from a superfluid to a
  Mott insulator in a gas of ultracold atoms}},}\ }\href {\doibase
  10.1038/415039a} {\bibfield  {journal} {\bibinfo  {journal} {Nature
  (London)}\ }\textbf {\bibinfo {volume} {415}},\ \bibinfo {pages} {39--44}
  (\bibinfo {year} {2002})}\BibitemShut {NoStop}%
\bibitem [{\citenamefont {Paredes}\ \emph {et~al.}(2004)\citenamefont
  {Paredes}, \citenamefont {Widera}, \citenamefont {Murg}, \citenamefont
  {Mandel}, \citenamefont {F{\"o}lling}, \citenamefont {Cirac}, \citenamefont
  {Shlyapnikov}, \citenamefont {H{\"a}nsch},\ and\ \citenamefont
  {Bloch}}]{Paredes:TonksGirardeauGas:N04}%
  \BibitemOpen
  \bibfield  {author} {\bibinfo {author} {\bibfnamefont {B.}~\bibnamefont
  {Paredes}}, \bibinfo {author} {\bibfnamefont {A.}~\bibnamefont {Widera}},
  \bibinfo {author} {\bibfnamefont {V.}~\bibnamefont {Murg}}, \bibinfo {author}
  {\bibfnamefont {O.}~\bibnamefont {Mandel}}, \bibinfo {author} {\bibfnamefont
  {S.}~\bibnamefont {F{\"o}lling}}, \bibinfo {author} {\bibfnamefont {J.~I.}\
  \bibnamefont {Cirac}}, \bibinfo {author} {\bibfnamefont {G.~V.}\ \bibnamefont
  {Shlyapnikov}}, \bibinfo {author} {\bibfnamefont {T.~W.}\ \bibnamefont
  {H{\"a}nsch}}, \ and\ \bibinfo {author} {\bibfnamefont {I.}~\bibnamefont
  {Bloch}},\ }\bibfield  {title} {\enquote {\bibinfo {title} {{Tonks-Girardeau
  gas of ultracold atoms in an optical lattice}},}\ }\href {\doibase
  10.1038/nature02530} {\bibfield  {journal} {\bibinfo  {journal} {Nature
  (London)}\ }\textbf {\bibinfo {volume} {429}},\ \bibinfo {pages} {277--281}
  (\bibinfo {year} {2004})}\BibitemShut {NoStop}%
\bibitem [{\citenamefont {Moore}\ \emph {et~al.}(1995)\citenamefont {Moore},
  \citenamefont {Robinson}, \citenamefont {Bharucha}, \citenamefont
  {Sundaram},\ and\ \citenamefont
  {Raizen}}]{Moore:AtomOpticsRealizationQKR:PRL95}%
  \BibitemOpen
  \bibfield  {author} {\bibinfo {author} {\bibfnamefont {F.~L.}\ \bibnamefont
  {Moore}}, \bibinfo {author} {\bibfnamefont {J.~C.}\ \bibnamefont {Robinson}},
  \bibinfo {author} {\bibfnamefont {C.~F.}\ \bibnamefont {Bharucha}}, \bibinfo
  {author} {\bibfnamefont {B.}~\bibnamefont {Sundaram}}, \ and\ \bibinfo
  {author} {\bibfnamefont {M.~G.}\ \bibnamefont {Raizen}},\ }\bibfield  {title}
  {\enquote {\bibinfo {title} {{Atom Optics Realization of the Quantum
  $\delta$-Kicked Rotor}},}\ }\href {\doibase 10.1103/PhysRevLett.75.4598}
  {\bibfield  {journal} {\bibinfo  {journal} {Phys. Rev. Lett.}\ }\textbf
  {\bibinfo {volume} {75}},\ \bibinfo {pages} {4598--4601} (\bibinfo {year}
  {1995})}\BibitemShut {NoStop}%
\bibitem [{\citenamefont {Billy}\ \emph {et~al.}(2008)\citenamefont {Billy},
  \citenamefont {Josse}, \citenamefont {Zuo}, \citenamefont {Bernard},
  \citenamefont {Hambrecht}, \citenamefont {Lugan}, \citenamefont
  {Cl{\'e}ment}, \citenamefont {Sanchez-Palencia}, \citenamefont {Bouyer},\
  and\ \citenamefont {Aspect}}]{Billy:AndersonBEC1D:N08}%
  \BibitemOpen
  \bibfield  {author} {\bibinfo {author} {\bibfnamefont {J.}~\bibnamefont
  {Billy}}, \bibinfo {author} {\bibfnamefont {V.}~\bibnamefont {Josse}},
  \bibinfo {author} {\bibfnamefont {Z.}~\bibnamefont {Zuo}}, \bibinfo {author}
  {\bibfnamefont {A.}~\bibnamefont {Bernard}}, \bibinfo {author} {\bibfnamefont
  {B.}~\bibnamefont {Hambrecht}}, \bibinfo {author} {\bibfnamefont
  {P.}~\bibnamefont {Lugan}}, \bibinfo {author} {\bibfnamefont
  {D.}~\bibnamefont {Cl{\'e}ment}}, \bibinfo {author} {\bibfnamefont
  {L.}~\bibnamefont {Sanchez-Palencia}}, \bibinfo {author} {\bibfnamefont
  {P.}~\bibnamefont {Bouyer}}, \ and\ \bibinfo {author} {\bibfnamefont
  {A.}~\bibnamefont {Aspect}},\ }\bibfield  {title} {\enquote {\bibinfo {title}
  {{Direct observation of Anderson localization of matter-waves in a controlled
  disorder}},}\ }\href {\doibase 10.1038/nature07000} {\bibfield  {journal}
  {\bibinfo  {journal} {Nature (London)}\ }\textbf {\bibinfo {volume} {453}},\
  \bibinfo {pages} {891--894} (\bibinfo {year} {2008})}\BibitemShut {NoStop}%
\bibitem [{\citenamefont {Roati}\ \emph {et~al.}(2008)\citenamefont {Roati},
  \citenamefont {d'Errico}, \citenamefont {Fallani}, \citenamefont {Fattori},
  \citenamefont {Fort}, \citenamefont {Zaccanti}, \citenamefont {Modugno},
  \citenamefont {Modugno},\ and\ \citenamefont
  {Inguscio}}]{Roati:AubryAndreBEC1D:N08}%
  \BibitemOpen
  \bibfield  {author} {\bibinfo {author} {\bibfnamefont {G.}~\bibnamefont
  {Roati}}, \bibinfo {author} {\bibfnamefont {C.}~\bibnamefont {d'Errico}},
  \bibinfo {author} {\bibfnamefont {L.}~\bibnamefont {Fallani}}, \bibinfo
  {author} {\bibfnamefont {M.}~\bibnamefont {Fattori}}, \bibinfo {author}
  {\bibfnamefont {C.}~\bibnamefont {Fort}}, \bibinfo {author} {\bibfnamefont
  {M.}~\bibnamefont {Zaccanti}}, \bibinfo {author} {\bibfnamefont
  {G.}~\bibnamefont {Modugno}}, \bibinfo {author} {\bibfnamefont
  {M.}~\bibnamefont {Modugno}}, \ and\ \bibinfo {author} {\bibfnamefont
  {M.}~\bibnamefont {Inguscio}},\ }\bibfield  {title} {\enquote {\bibinfo
  {title} {{Anderson localization of a non-interacting Bose-Einstein
  condensate}},}\ }\href {\doibase 10.1038/nature07071} {\bibfield  {journal}
  {\bibinfo  {journal} {Nature (London)}\ }\textbf {\bibinfo {volume} {453}},\
  \bibinfo {pages} {895--898} (\bibinfo {year} {2008})}\BibitemShut {NoStop}%
\bibitem [{\citenamefont {Jendrzejewski}\ \emph {et~al.}(2012)\citenamefont
  {Jendrzejewski}, \citenamefont {Bernard}, \citenamefont {M{\"u}ller},
  \citenamefont {Cheinet}, \citenamefont {Josse}, \citenamefont {Piraud},
  \citenamefont {Pezz{\`e}}, \citenamefont {Sanchez-Palencia}, \citenamefont
  {Aspect},\ and\ \citenamefont {Bouyer}}]{Jendrzejewski:AndersonLoc3D:NP12}%
  \BibitemOpen
  \bibfield  {author} {\bibinfo {author} {\bibfnamefont {F.}~\bibnamefont
  {Jendrzejewski}}, \bibinfo {author} {\bibfnamefont {A.}~\bibnamefont
  {Bernard}}, \bibinfo {author} {\bibfnamefont {K.}~\bibnamefont {M{\"u}ller}},
  \bibinfo {author} {\bibfnamefont {P.}~\bibnamefont {Cheinet}}, \bibinfo
  {author} {\bibfnamefont {V.}~\bibnamefont {Josse}}, \bibinfo {author}
  {\bibfnamefont {M.}~\bibnamefont {Piraud}}, \bibinfo {author} {\bibfnamefont
  {L.}~\bibnamefont {Pezz{\`e}}}, \bibinfo {author} {\bibfnamefont
  {L.}~\bibnamefont {Sanchez-Palencia}}, \bibinfo {author} {\bibfnamefont
  {A.}~\bibnamefont {Aspect}}, \ and\ \bibinfo {author} {\bibfnamefont
  {P.}~\bibnamefont {Bouyer}},\ }\bibfield  {title} {\enquote {\bibinfo {title}
  {{Three-dimensional localization of ultracold atoms in an optical disordered
  potential}},}\ }\href {\doibase 10.1038/nphys2256} {\bibfield  {journal}
  {\bibinfo  {journal} {Nat. Phys.}\ }\textbf {\bibinfo {volume} {8}},\
  \bibinfo {pages} {398--403} (\bibinfo {year} {2012})}\BibitemShut {NoStop}%
\bibitem [{\citenamefont {Semeghini}\ \emph {et~al.}(2015)\citenamefont
  {Semeghini}, \citenamefont {Landini}, \citenamefont {Castilho}, \citenamefont
  {Roy}, \citenamefont {Spagnolli}, \citenamefont {Trenkwalder}, \citenamefont
  {Fattori}, \citenamefont {Inguscio},\ and\ \citenamefont
  {Modugno}}]{Semeghini:MobilityEdgeAnderson:NP15}%
  \BibitemOpen
  \bibfield  {author} {\bibinfo {author} {\bibfnamefont {G.}~\bibnamefont
  {Semeghini}}, \bibinfo {author} {\bibfnamefont {M.}~\bibnamefont {Landini}},
  \bibinfo {author} {\bibfnamefont {P.}~\bibnamefont {Castilho}}, \bibinfo
  {author} {\bibfnamefont {S.}~\bibnamefont {Roy}}, \bibinfo {author}
  {\bibfnamefont {G.}~\bibnamefont {Spagnolli}}, \bibinfo {author}
  {\bibfnamefont {A.}~\bibnamefont {Trenkwalder}}, \bibinfo {author}
  {\bibfnamefont {M.}~\bibnamefont {Fattori}}, \bibinfo {author} {\bibfnamefont
  {M.}~\bibnamefont {Inguscio}}, \ and\ \bibinfo {author} {\bibfnamefont
  {G.}~\bibnamefont {Modugno}},\ }\bibfield  {title} {\enquote {\bibinfo
  {title} {{Measurement of the mobility edge for 3D Anderson localization}},}\
  }\href {\doibase 10.1038/nphys3339} {\bibfield  {journal} {\bibinfo
  {journal} {Nat. Phys.}\ }\textbf {\bibinfo {volume} {11}},\ \bibinfo {pages}
  {554--559} (\bibinfo {year} {2015})}\BibitemShut {NoStop}%
\bibitem [{\citenamefont {Chab{\'e}}\ \emph {et~al.}(2008)\citenamefont
  {Chab{\'e}}, \citenamefont {Lemari{\'e}}, \citenamefont {Gr{\'e}maud},
  \citenamefont {Delande}, \citenamefont {Szriftgiser},\ and\ \citenamefont
  {Garreau}}]{Chabe:Anderson:PRL08}%
  \BibitemOpen
  \bibfield  {author} {\bibinfo {author} {\bibfnamefont {J.}~\bibnamefont
  {Chab{\'e}}}, \bibinfo {author} {\bibfnamefont {G.}~\bibnamefont
  {Lemari{\'e}}}, \bibinfo {author} {\bibfnamefont {B.}~\bibnamefont
  {Gr{\'e}maud}}, \bibinfo {author} {\bibfnamefont {D.}~\bibnamefont
  {Delande}}, \bibinfo {author} {\bibfnamefont {P.}~\bibnamefont
  {Szriftgiser}}, \ and\ \bibinfo {author} {\bibfnamefont {J.~C.}\ \bibnamefont
  {Garreau}},\ }\bibfield  {title} {\enquote {\bibinfo {title} {{Experimental
  Observation of the Anderson Metal-Insulator Transition with Atomic Matter
  Waves}},}\ }\href {\doibase 10.1103/PhysRevLett.101.255702} {\bibfield
  {journal} {\bibinfo  {journal} {Phys. Rev. Lett.}\ }\textbf {\bibinfo
  {volume} {101}},\ \bibinfo {pages} {255702} (\bibinfo {year}
  {2008})}\BibitemShut {NoStop}%
\bibitem [{\citenamefont {Lopez}\ \emph {et~al.}(2012)\citenamefont {Lopez},
  \citenamefont {Cl{\'e}ment}, \citenamefont {Szriftgiser}, \citenamefont
  {Garreau},\ and\ \citenamefont
  {Delande}}]{Lopez:ExperimentalTestOfUniversality:PRL12}%
  \BibitemOpen
  \bibfield  {author} {\bibinfo {author} {\bibfnamefont {M.}~\bibnamefont
  {Lopez}}, \bibinfo {author} {\bibfnamefont {J.-F.}\ \bibnamefont
  {Cl{\'e}ment}}, \bibinfo {author} {\bibfnamefont {P.}~\bibnamefont
  {Szriftgiser}}, \bibinfo {author} {\bibfnamefont {J.~C.}\ \bibnamefont
  {Garreau}}, \ and\ \bibinfo {author} {\bibfnamefont {D.}~\bibnamefont
  {Delande}},\ }\bibfield  {title} {\enquote {\bibinfo {title} {{Experimental
  Test of Universality of the Anderson Transition}},}\ }\href {\doibase
  10.1103/PhysRevLett.108.095701} {\bibfield  {journal} {\bibinfo  {journal}
  {Phys. Rev. Lett.}\ }\textbf {\bibinfo {volume} {108}},\ \bibinfo {pages}
  {095701} (\bibinfo {year} {2012})}\BibitemShut {NoStop}%
\bibitem [{\citenamefont {Manai}\ \emph {et~al.}(2015)\citenamefont {Manai},
  \citenamefont {Cl{\'e}ment}, \citenamefont {Chicireanu}, \citenamefont
  {Hainaut}, \citenamefont {Garreau}, \citenamefont {Szriftgiser},\ and\
  \citenamefont {Delande}}]{Manai:Anderson2DKR:PRL15}%
  \BibitemOpen
  \bibfield  {author} {\bibinfo {author} {\bibfnamefont {I.}~\bibnamefont
  {Manai}}, \bibinfo {author} {\bibfnamefont {J.-F.}\ \bibnamefont
  {Cl{\'e}ment}}, \bibinfo {author} {\bibfnamefont {R.}~\bibnamefont
  {Chicireanu}}, \bibinfo {author} {\bibfnamefont {C.}~\bibnamefont {Hainaut}},
  \bibinfo {author} {\bibfnamefont {J.~C.}\ \bibnamefont {Garreau}}, \bibinfo
  {author} {\bibfnamefont {P.}~\bibnamefont {Szriftgiser}}, \ and\ \bibinfo
  {author} {\bibfnamefont {D.}~\bibnamefont {Delande}},\ }\bibfield  {title}
  {\enquote {\bibinfo {title} {{Experimental Observation of Two-Dimensional
  Anderson Localization with the Atomic Kicked Rotor}},}\ }\href {\doibase
  10.1103/PhysRevLett.115.240603} {\bibfield  {journal} {\bibinfo  {journal}
  {Phys. Rev. Lett.}\ }\textbf {\bibinfo {volume} {115}},\ \bibinfo {pages}
  {240603} (\bibinfo {year} {2015})}\BibitemShut {NoStop}%
\bibitem [{\citenamefont {Hainaut}\ \emph
  {et~al.}(2018{\natexlab{a}})\citenamefont {Hainaut}, \citenamefont {Manai},
  \citenamefont {Cl{\'e}ment}, \citenamefont {Garreau}, \citenamefont
  {Szriftgiser}, \citenamefont {Lemari{\'e}}, \citenamefont {Cherroret},
  \citenamefont {Delande},\ and\ \citenamefont
  {Chicireanu}}]{Hainaut:CFS:NCM18}%
  \BibitemOpen
  \bibfield  {author} {\bibinfo {author} {\bibfnamefont {C.}~\bibnamefont
  {Hainaut}}, \bibinfo {author} {\bibfnamefont {I.}~\bibnamefont {Manai}},
  \bibinfo {author} {\bibfnamefont {J.-F.}\ \bibnamefont {Cl{\'e}ment}},
  \bibinfo {author} {\bibfnamefont {J.~C.}\ \bibnamefont {Garreau}}, \bibinfo
  {author} {\bibfnamefont {P.}~\bibnamefont {Szriftgiser}}, \bibinfo {author}
  {\bibfnamefont {G.}~\bibnamefont {Lemari{\'e}}}, \bibinfo {author}
  {\bibfnamefont {N.}~\bibnamefont {Cherroret}}, \bibinfo {author}
  {\bibfnamefont {D.}~\bibnamefont {Delande}}, \ and\ \bibinfo {author}
  {\bibfnamefont {R.}~\bibnamefont {Chicireanu}},\ }\bibfield  {title}
  {\enquote {\bibinfo {title} {{Controlling symmetry and localization with an
  artificial gauge field in a disordered quantum system}},}\ }\href {\doibase
  10.1038/s41467-018-03481-9} {\bibfield  {journal} {\bibinfo  {journal} {Nat.
  Commun.}\ }\textbf {\bibinfo {volume} {9}},\ \bibinfo {pages} {1382}
  (\bibinfo {year} {2018}{\natexlab{a}})}\BibitemShut {NoStop}%
\bibitem [{\citenamefont
  {Garreau}(2017)}]{Garreau:QuantumSimulationOfDisordered:CRP17}%
  \BibitemOpen
  \bibfield  {author} {\bibinfo {author} {\bibfnamefont {J.~C.}\ \bibnamefont
  {Garreau}},\ }\bibfield  {title} {\enquote {\bibinfo {title} {{Quantum
  simulation of disordered systems with cold atoms}},}\ }\href {\doibase
  http://dx.doi.org/10.1016/j.crhy.2016.09.002} {\bibfield  {journal} {\bibinfo
   {journal} {Compt. Rendus Phys.}\ }\textbf {\bibinfo {volume} {18}},\
  \bibinfo {pages} {31 -- 46} (\bibinfo {year} {2017})}\BibitemShut {NoStop}%
\bibitem [{\citenamefont {Hainaut}\ \emph
  {et~al.}(2018{\natexlab{b}})\citenamefont {Hainaut}, \citenamefont {Rançon},
  \citenamefont {Cl{\'e}ment}, \citenamefont {Garreau}, \citenamefont
  {Szriftgiser}, \citenamefont {Chicireanu},\ and\ \citenamefont
  {Delande}}]{Hainaut:RatchetEffectQKR:PRA18}%
  \BibitemOpen
  \bibfield  {author} {\bibinfo {author} {\bibfnamefont {C.}~\bibnamefont
  {Hainaut}}, \bibinfo {author} {\bibfnamefont {A.}~\bibnamefont {Ran\c con}},
  \bibinfo {author} {\bibfnamefont {J.~F.}\ \bibnamefont {Cl{\'e}ment}},
  \bibinfo {author} {\bibfnamefont {J.~C.}\ \bibnamefont {Garreau}}, \bibinfo
  {author} {\bibfnamefont {P.}~\bibnamefont {Szriftgiser}}, \bibinfo {author}
  {\bibfnamefont {R.}~\bibnamefont {Chicireanu}}, \ and\ \bibinfo {author}
  {\bibfnamefont {D.}~\bibnamefont {Delande}},\ }\bibfield  {title} {\enquote
  {\bibinfo {title} {{Ratchet effect in the quantum kicked rotor and its
  destruction by dynamical localization}},}\ }\href {\doibase
  10.1103/PhysRevA.97.061601} {\bibfield  {journal} {\bibinfo  {journal} {Phys.
  Rev. A}\ }\textbf {\bibinfo {volume} {97}},\ \bibinfo {pages} {061601}
  (\bibinfo {year} {2018}{\natexlab{b}})}\BibitemShut {NoStop}%
\bibitem [{\citenamefont {Abrahams}\ \emph {et~al.}(1979)\citenamefont
  {Abrahams}, \citenamefont {Anderson}, \citenamefont {Licciardello},\ and\
  \citenamefont {Ramakrishnan}}]{Abrahams:Scaling:PRL79}%
  \BibitemOpen
  \bibfield  {author} {\bibinfo {author} {\bibfnamefont {E.}~\bibnamefont
  {Abrahams}}, \bibinfo {author} {\bibfnamefont {P.~W.}\ \bibnamefont
  {Anderson}}, \bibinfo {author} {\bibfnamefont {D.~C.}\ \bibnamefont
  {Licciardello}}, \ and\ \bibinfo {author} {\bibfnamefont {T.~V.}\
  \bibnamefont {Ramakrishnan}},\ }\bibfield  {title} {\enquote {\bibinfo
  {title} {{Scaling Theory of Localization\string: Absence of Quantum Diffusion
  in Two Dimensions}},}\ }\href {\doibase 10.1103/PhysRevLett.42.673}
  {\bibfield  {journal} {\bibinfo  {journal} {Phys. Rev. Lett.}\ }\textbf
  {\bibinfo {volume} {42}},\ \bibinfo {pages} {673--676} (\bibinfo {year}
  {1979})}\BibitemShut {NoStop}%
\bibitem [{\citenamefont {Deng}\ \emph {et~al.}(1999)\citenamefont {Deng},
  \citenamefont {Hagley}, \citenamefont {Denschlag}, \citenamefont {Simsarian},
  \citenamefont {Edwards}, \citenamefont {Clark}, \citenamefont {Helmerson},
  \citenamefont {Rolston},\ and\ \citenamefont
  {Phillips}}]{Philips:QuantumResTalbot:PRL99}%
  \BibitemOpen
  \bibfield  {author} {\bibinfo {author} {\bibfnamefont {L.}~\bibnamefont
  {Deng}}, \bibinfo {author} {\bibfnamefont {E.~W.}\ \bibnamefont {Hagley}},
  \bibinfo {author} {\bibfnamefont {J.}~\bibnamefont {Denschlag}}, \bibinfo
  {author} {\bibfnamefont {J.~E.}\ \bibnamefont {Simsarian}}, \bibinfo {author}
  {\bibfnamefont {M.}~\bibnamefont {Edwards}}, \bibinfo {author} {\bibfnamefont
  {C.~W.}\ \bibnamefont {Clark}}, \bibinfo {author} {\bibfnamefont
  {K.}~\bibnamefont {Helmerson}}, \bibinfo {author} {\bibfnamefont {S.~L.}\
  \bibnamefont {Rolston}}, \ and\ \bibinfo {author} {\bibfnamefont {W.~D.}\
  \bibnamefont {Phillips}},\ }\bibfield  {title} {\enquote {\bibinfo {title}
  {{Temporal, Matter-Wave-Dispersion Talbot Effect}},}\ }\href {\doibase
  10.1103/PhysRevLett.83.5407} {\bibfield  {journal} {\bibinfo  {journal}
  {Phys. Rev. Lett.}\ }\textbf {\bibinfo {volume} {83}},\ \bibinfo {pages}
  {5407--5411} (\bibinfo {year} {1999})}\BibitemShut {NoStop}%
\bibitem [{\citenamefont {Dana}\ \emph {et~al.}(2008)\citenamefont {Dana},
  \citenamefont {Ramareddy}, \citenamefont {Talukdar},\ and\ \citenamefont
  {Summy}}]{Summy:QRRarchets:PRL08}%
  \BibitemOpen
  \bibfield  {author} {\bibinfo {author} {\bibfnamefont {I.}~\bibnamefont
  {Dana}}, \bibinfo {author} {\bibfnamefont {V.}~\bibnamefont {Ramareddy}},
  \bibinfo {author} {\bibfnamefont {I.}~\bibnamefont {Talukdar}}, \ and\
  \bibinfo {author} {\bibfnamefont {G.~S.}\ \bibnamefont {Summy}},\ }\bibfield
  {title} {\enquote {\bibinfo {title} {{Experimental Realization of
  Quantum-Resonance Ratchets at Arbitrary Quasimomenta}},}\ }\href {\doibase
  10.1103/PhysRevLett.100.024103} {\bibfield  {journal} {\bibinfo  {journal}
  {Phys. Rev. Lett.}\ }\textbf {\bibinfo {volume} {100}},\ \bibinfo {pages}
  {024103} (\bibinfo {year} {2008})}\BibitemShut {NoStop}%
\bibitem [{\citenamefont {Hensinger}\ \emph {et~al.}(2001)\citenamefont
  {Hensinger}, \citenamefont {Haeffner}, \citenamefont {Browaeys},
  \citenamefont {Heckenberg}, \citenamefont {Helmerson}, \citenamefont
  {McKenzie}, \citenamefont {Milburn}, \citenamefont {Phillips}, \citenamefont
  {Rolston}, \citenamefont {Rubinsztein-Dunlop},\ and\ \citenamefont
  {Upcroft}}]{Rubins-Dunlop:ChaosAssitTunnel:Nature01}%
  \BibitemOpen
  \bibfield  {author} {\bibinfo {author} {\bibfnamefont {W.~K.}\ \bibnamefont
  {Hensinger}}, \bibinfo {author} {\bibfnamefont {H.}~\bibnamefont {Haeffner}},
  \bibinfo {author} {\bibfnamefont {A.}~\bibnamefont {Browaeys}}, \bibinfo
  {author} {\bibfnamefont {N.~R.}\ \bibnamefont {Heckenberg}}, \bibinfo
  {author} {\bibfnamefont {K.}~\bibnamefont {Helmerson}}, \bibinfo {author}
  {\bibfnamefont {C.}~\bibnamefont {McKenzie}}, \bibinfo {author}
  {\bibfnamefont {G.~J.}\ \bibnamefont {Milburn}}, \bibinfo {author}
  {\bibfnamefont {W.~D.}\ \bibnamefont {Phillips}}, \bibinfo {author}
  {\bibfnamefont {S.~L.}\ \bibnamefont {Rolston}}, \bibinfo {author}
  {\bibfnamefont {H.}~\bibnamefont {Rubinsztein-Dunlop}}, \ and\ \bibinfo
  {author} {\bibfnamefont {B.}~\bibnamefont {Upcroft}},\ }\bibfield  {title}
  {\enquote {\bibinfo {title} {{Dynamical tunneling of ultracold atoms}},}\
  }\href@noop {} {\bibfield  {journal} {\bibinfo  {journal} {Nature (London)}\
  }\textbf {\bibinfo {volume} {412}},\ \bibinfo {pages} {52--55} (\bibinfo
  {year} {2001})}\BibitemShut {NoStop}%
\bibitem [{\citenamefont {Steck}\ \emph {et~al.}(2001)\citenamefont {Steck},
  \citenamefont {Oskay},\ and\ \citenamefont
  {Raizen}}]{Raizen:ChaosAssistTunnel:Science01}%
  \BibitemOpen
  \bibfield  {author} {\bibinfo {author} {\bibfnamefont {D.~A.}\ \bibnamefont
  {Steck}}, \bibinfo {author} {\bibfnamefont {W.~H.}\ \bibnamefont {Oskay}}, \
  and\ \bibinfo {author} {\bibfnamefont {M.~G.}\ \bibnamefont {Raizen}},\
  }\bibfield  {title} {\enquote {\bibinfo {title} {{Observation of
  chaos-assisted tunneling between islands of stability}},}\ }\href@noop {}
  {\bibfield  {journal} {\bibinfo  {journal} {Science}\ }\textbf {\bibinfo
  {volume} {293}},\ \bibinfo {pages} {274--278} (\bibinfo {year}
  {2001})}\BibitemShut {NoStop}%
\bibitem [{\citenamefont {Sadgrove}\ \emph {et~al.}(2005)\citenamefont
  {Sadgrove}, \citenamefont {Wimberger}, \citenamefont {Parkins},\ and\
  \citenamefont {Leonhardt}}]{Leonhardt:TransportKR:PRL05}%
  \BibitemOpen
  \bibfield  {author} {\bibinfo {author} {\bibfnamefont {M.}~\bibnamefont
  {Sadgrove}}, \bibinfo {author} {\bibfnamefont {S.}~\bibnamefont {Wimberger}},
  \bibinfo {author} {\bibfnamefont {S.}~\bibnamefont {Parkins}}, \ and\
  \bibinfo {author} {\bibfnamefont {R.}~\bibnamefont {Leonhardt}},\ }\bibfield
  {title} {\enquote {\bibinfo {title} {{Ballistic and Localized Transport for
  the Atom Optics Kicked Rotor in the Limit of a Vanishing Kicking Period}},}\
  }\href {\doibase 10.1103/PhysRevLett.94.174103} {\bibfield  {journal}
  {\bibinfo  {journal} {Phys. Rev. Lett.}\ }\textbf {\bibinfo {volume} {94}},\
  \bibinfo {pages} {174103} (\bibinfo {year} {2005})}\BibitemShut {NoStop}%
\bibitem [{\citenamefont {Wimberger}\ \emph {et~al.}(2005)\citenamefont
  {Wimberger}, \citenamefont {Sadgrove}, \citenamefont {Parkins},\ and\
  \citenamefont {Leonhardt}}]{Wimberger:QRScalingLaw:PRA05}%
  \BibitemOpen
  \bibfield  {author} {\bibinfo {author} {\bibfnamefont {S.}~\bibnamefont
  {Wimberger}}, \bibinfo {author} {\bibfnamefont {M.}~\bibnamefont {Sadgrove}},
  \bibinfo {author} {\bibfnamefont {S.}~\bibnamefont {Parkins}}, \ and\
  \bibinfo {author} {\bibfnamefont {R.}~\bibnamefont {Leonhardt}},\ }\bibfield
  {title} {\enquote {\bibinfo {title} {{Experimental verification of a
  one-parameter scaling law for the quantum and "classical" resonances of the
  atom-optics kicked rotor}},}\ }\href {\doibase 10.1103/PhysRevA.71.053404}
  {\bibfield  {journal} {\bibinfo  {journal} {Phys. Rev. A}\ }\textbf {\bibinfo
  {volume} {71}},\ \bibinfo {pages} {053404} (\bibinfo {year}
  {2005})}\BibitemShut {NoStop}%
\bibitem [{\citenamefont {Ma}\ \emph {et~al.}(2004)\citenamefont {Ma},
  \citenamefont {d'Arcy},\ and\ \citenamefont
  {Gardiner}}]{DArcy:GravityQuantRes:PRL04}%
  \BibitemOpen
  \bibfield  {author} {\bibinfo {author} {\bibfnamefont {Z.~Y.}\ \bibnamefont
  {Ma}}, \bibinfo {author} {\bibfnamefont {M.~B.}\ \bibnamefont {d'Arcy}}, \
  and\ \bibinfo {author} {\bibfnamefont {S.~A.}\ \bibnamefont {Gardiner}},\
  }\bibfield  {title} {\enquote {\bibinfo {title} {{Gravity-Sensitive Quantum
  Dynamics in Cold Atoms}},}\ }\href {\doibase 10.1103/PhysRevLett.93.164101}
  {\bibfield  {journal} {\bibinfo  {journal} {Phys. Rev. Lett.}\ }\textbf
  {\bibinfo {volume} {93}},\ \bibinfo {pages} {164101} (\bibinfo {year}
  {2004})}\BibitemShut {NoStop}%
\bibitem [{\citenamefont {Casati}\ \emph {et~al.}(1979)\citenamefont {Casati},
  \citenamefont {Chirikov}, \citenamefont {Ford},\ and\ \citenamefont
  {Izrailev}}]{Casati:LocDynFirst:LNP79}%
  \BibitemOpen
  \bibfield  {author} {\bibinfo {author} {\bibfnamefont {G.}~\bibnamefont
  {Casati}}, \bibinfo {author} {\bibfnamefont {B.~V.}\ \bibnamefont
  {Chirikov}}, \bibinfo {author} {\bibfnamefont {J.}~\bibnamefont {Ford}}, \
  and\ \bibinfo {author} {\bibfnamefont {F.~M.}\ \bibnamefont {Izrailev}},\
  }\enquote {\bibinfo {title} {{Stochastic behavior of a quantum pendulum under
  periodic perturbation}},}\ in\ \href {\doibase 10.1007/BFb0021757} {\emph
  {\bibinfo {booktitle} {{Stochastic Behavior in Classical and Quantum
  Systems}}}},\ Vol.~\bibinfo {volume} {93},\ \bibinfo {editor} {edited by\
  \bibinfo {editor} {\bibnamefont {{G. Casati and J. Ford}}}}\ (\bibinfo
  {publisher} {{Springer-Verlag}},\ \bibinfo {address} {{Berlin, Germany}},\
  \bibinfo {year} {1979})\ pp.\ \bibinfo {pages} {334--352}\BibitemShut
  {NoStop}%
\bibitem [{\citenamefont {Grempel}\ \emph {et~al.}(1984)\citenamefont
  {Grempel}, \citenamefont {Prange},\ and\ \citenamefont
  {Fishman}}]{Fishman:LocDynAnderson:PRA84}%
  \BibitemOpen
  \bibfield  {author} {\bibinfo {author} {\bibfnamefont {D.~R.}\ \bibnamefont
  {Grempel}}, \bibinfo {author} {\bibfnamefont {R.~E.}\ \bibnamefont {Prange}},
  \ and\ \bibinfo {author} {\bibfnamefont {S.}~\bibnamefont {Fishman}},\
  }\bibfield  {title} {\enquote {\bibinfo {title} {{Quantum dynamics of a
  nonintegrable system}},}\ }\href {\doibase 10.1103/PhysRevA.29.1639}
  {\bibfield  {journal} {\bibinfo  {journal} {Phys. Rev. A}\ }\textbf {\bibinfo
  {volume} {29}},\ \bibinfo {pages} {1639--1647} (\bibinfo {year}
  {1984})}\BibitemShut {NoStop}%
\bibitem [{Note1()}]{Note1}%
  \BibitemOpen
  \bibinfo {note} {The kinetic energy term can be neglected compared to the
  $\delta $ function in the kick term.}\BibitemShut {Stop}%
\bibitem [{\citenamefont {Fishman}\ \emph {et~al.}(1982)\citenamefont
  {Fishman}, \citenamefont {Grempel},\ and\ \citenamefont
  {Prange}}]{Fishman:LocDynAnders:PRL82}%
  \BibitemOpen
  \bibfield  {author} {\bibinfo {author} {\bibfnamefont {S.}~\bibnamefont
  {Fishman}}, \bibinfo {author} {\bibfnamefont {D.~R.}\ \bibnamefont
  {Grempel}}, \ and\ \bibinfo {author} {\bibfnamefont {R.~E.}\ \bibnamefont
  {Prange}},\ }\bibfield  {title} {\enquote {\bibinfo {title} {{Chaos, Quantum
  Recurrences, and Anderson Localization}},}\ }\href {\doibase
  10.1103/PhysRevLett.49.509} {\bibfield  {journal} {\bibinfo  {journal} {Phys.
  Rev. Lett.}\ }\textbf {\bibinfo {volume} {49}},\ \bibinfo {pages} {509--512}
  (\bibinfo {year} {1982})}\BibitemShut {NoStop}%
\bibitem [{Note2()}]{Note2}%
  \BibitemOpen
  \bibinfo {note} {See e.g. \protect \cite
  {Garreau:QuantumSimulationOfDisordered:CRP17} and references
  therein.}\BibitemShut {Stop}%
\bibitem [{\citenamefont {Shepelyansky}(1987)}]{Shepelyansky:Bicolor:PD87}%
  \BibitemOpen
  \bibfield  {author} {\bibinfo {author} {\bibfnamefont {D.~L.}\ \bibnamefont
  {Shepelyansky}},\ }\bibfield  {title} {\enquote {\bibinfo {title}
  {{Localization of diffusive excitation in multi-level systems}},}\ }\href
  {\doibase http://dx.doi.org/10.1016/0167-2789(87)90123-0} {\bibfield
  {journal} {\bibinfo  {journal} {Physica D}\ }\textbf {\bibinfo {volume}
  {28}},\ \bibinfo {pages} {103--114} (\bibinfo {year} {1987})}\BibitemShut
  {NoStop}%
\bibitem [{\citenamefont {Cherroret}\ \emph {et~al.}(2014)\citenamefont
  {Cherroret}, \citenamefont {Vermersch}, \citenamefont {Garreau},\ and\
  \citenamefont {Delande}}]{Cherroret:AndersonNonlinearInteractions:PRL14}%
  \BibitemOpen
  \bibfield  {author} {\bibinfo {author} {\bibfnamefont {N.}~\bibnamefont
  {Cherroret}}, \bibinfo {author} {\bibfnamefont {B.}~\bibnamefont
  {Vermersch}}, \bibinfo {author} {\bibfnamefont {J.~C.}\ \bibnamefont
  {Garreau}}, \ and\ \bibinfo {author} {\bibfnamefont {D.}~\bibnamefont
  {Delande}},\ }\bibfield  {title} {\enquote {\bibinfo {title} {{How Nonlinear
  Interactions Challenge the Three-Dimensional Anderson Transition}},}\ }\href
  {\doibase 10.1103/PhysRevLett.112.170603} {\bibfield  {journal} {\bibinfo
  {journal} {Phys. Rev. Lett.}\ }\textbf {\bibinfo {volume} {112}},\ \bibinfo
  {pages} {170603} (\bibinfo {year} {2014})}\BibitemShut {NoStop}%
\bibitem [{Note3()}]{Note3}%
  \BibitemOpen
  \bibinfo {note} {In the RKR, ${\mathchar '26\mkern -9mu k}$ is not defined,
  as the free propagation phases are random, the pertinent parameter that we
  call $K$ for simplicity is equivalent to $K/{\mathchar '26\mkern -9mu k}$ for
  the standard QKR.}\BibitemShut {Stop}%
\bibitem [{\citenamefont {White}\ \emph {et~al.}(2013)\citenamefont {White},
  \citenamefont {Ruddell},\ and\ \citenamefont
  {Hoogerland}}]{White:ExperimentalQuantumRatchet:PRA13}%
  \BibitemOpen
  \bibfield  {author} {\bibinfo {author} {\bibfnamefont {D.~H.}\ \bibnamefont
  {White}}, \bibinfo {author} {\bibfnamefont {S.~K.}\ \bibnamefont {Ruddell}},
  \ and\ \bibinfo {author} {\bibfnamefont {M.~D.}\ \bibnamefont {Hoogerland}},\
  }\bibfield  {title} {\enquote {\bibinfo {title} {{Experimental realization of
  a quantum ratchet through phase modulation}},}\ }\href {\doibase
  10.1103/PhysRevA.88.063603} {\bibfield  {journal} {\bibinfo  {journal} {Phys.
  Rev. A}\ }\textbf {\bibinfo {volume} {88}},\ \bibinfo {pages} {063603}
  (\bibinfo {year} {2013})}\BibitemShut {NoStop}%
\bibitem [{\citenamefont {Rechester}\ \emph {et~al.}(1981)\citenamefont
  {Rechester}, \citenamefont {Rosenbluth},\ and\ \citenamefont
  {White}}]{Rechester:KRDiffCoeff:PRA81}%
  \BibitemOpen
  \bibfield  {author} {\bibinfo {author} {\bibfnamefont {A.~B.}\ \bibnamefont
  {Rechester}}, \bibinfo {author} {\bibfnamefont {M.~N.}\ \bibnamefont
  {Rosenbluth}}, \ and\ \bibinfo {author} {\bibfnamefont {R.~B.}\ \bibnamefont
  {White}},\ }\bibfield  {title} {\enquote {\bibinfo {title} {{Fourier-space
  paths applied to the calculation of diffusion for the Chirikov-Taylor
  model}},}\ }\href {\doibase 10.1103/PhysRevA.23.2664} {\bibfield  {journal}
  {\bibinfo  {journal} {Phys. Rev. A}\ }\textbf {\bibinfo {volume} {23}},\
  \bibinfo {pages} {2664--2672} (\bibinfo {year} {1981})}\BibitemShut {NoStop}%
\bibitem [{\citenamefont {Rechester}\ and\ \citenamefont
  {White}(1980)}]{Rechester:TurbulentDiffChirikovTaylor:PRL80}%
  \BibitemOpen
  \bibfield  {author} {\bibinfo {author} {\bibfnamefont {A.~B.}\ \bibnamefont
  {Rechester}}\ and\ \bibinfo {author} {\bibfnamefont {R.~B.}\ \bibnamefont
  {White}},\ }\bibfield  {title} {\enquote {\bibinfo {title} {{Calculation of
  Turbulent Diffusion for the Chirikov-Taylor Model}},}\ }\href {\doibase
  10.1103/PhysRevLett.44.1586} {\bibfield  {journal} {\bibinfo  {journal}
  {Phys. Rev. Lett.}\ }\textbf {\bibinfo {volume} {44}},\ \bibinfo {pages}
  {1586--1589} (\bibinfo {year} {1980})}\BibitemShut {NoStop}%
\bibitem [{\citenamefont {Klappauf}\ \emph {et~al.}(1998)\citenamefont
  {Klappauf}, \citenamefont {Oskay}, \citenamefont {Steck},\ and\ \citenamefont
  {Raizen}}]{Raizen:KRClassRes:PRL98}%
  \BibitemOpen
  \bibfield  {author} {\bibinfo {author} {\bibfnamefont {B.~G.}\ \bibnamefont
  {Klappauf}}, \bibinfo {author} {\bibfnamefont {W.~H.}\ \bibnamefont {Oskay}},
  \bibinfo {author} {\bibfnamefont {D.~A.}\ \bibnamefont {Steck}}, \ and\
  \bibinfo {author} {\bibfnamefont {M.~G.}\ \bibnamefont {Raizen}},\ }\bibfield
   {title} {\enquote {\bibinfo {title} {{Experimental Study of Quantum Dynamics
  in a Regime of Classical Anomalous Diffusion}},}\ }\href {\doibase
  10.1103/PhysRevLett.81.4044} {\bibfield  {journal} {\bibinfo  {journal}
  {Phys. Rev. Lett.}\ }\textbf {\bibinfo {volume} {81}},\ \bibinfo {pages}
  {4044--4047} (\bibinfo {year} {1998})}\BibitemShut {NoStop}%
\bibitem [{\citenamefont {Tian}\ \emph {et~al.}(2005)\citenamefont {Tian},
  \citenamefont {Kamenev},\ and\ \citenamefont
  {Larkin}}]{Tian:EhrenfestTimeDynamicalLoc:PRB05}%
  \BibitemOpen
  \bibfield  {author} {\bibinfo {author} {\bibfnamefont {C.}~\bibnamefont
  {Tian}}, \bibinfo {author} {\bibfnamefont {A.}~\bibnamefont {Kamenev}}, \
  and\ \bibinfo {author} {\bibfnamefont {A.}~\bibnamefont {Larkin}},\
  }\bibfield  {title} {\enquote {\bibinfo {title} {{Ehrenfest time in the weak
  dynamical localization}},}\ }\href {\doibase 10.1103/PhysRevB.72.045108}
  {\bibfield  {journal} {\bibinfo  {journal} {Phys. Rev. B}\ }\textbf {\bibinfo
  {volume} {72}},\ \bibinfo {pages} {045108} (\bibinfo {year}
  {2005})}\BibitemShut {NoStop}%
\bibitem [{Note4()}]{Note4}%
  \BibitemOpen
  \bibinfo {note} {The realization of a PSQKR in the unitary symmetry class is
  demonstrated in Ref.~\cite {Hainaut:CFS:NCM18}}\BibitemShut {NoStop}%
\bibitem [{\citenamefont {Lobkis}\ and\ \citenamefont
  {Weaver}(2005)}]{LobkisWeaver:SelfConsistentTransportLocWaves:PRE05}%
  \BibitemOpen
  \bibfield  {author} {\bibinfo {author} {\bibfnamefont {O.~I.}\ \bibnamefont
  {Lobkis}}\ and\ \bibinfo {author} {\bibfnamefont {R.~L.}\ \bibnamefont
  {Weaver}},\ }\bibfield  {title} {\enquote {\bibinfo {title} {{Self-consistent
  transport dynamics for localized waves}},}\ }\href {\doibase
  10.1103/PhysRevE.71.011112} {\bibfield  {journal} {\bibinfo  {journal} {Phys.
  Rev. E}\ }\textbf {\bibinfo {volume} {71}},\ \bibinfo {pages} {011112}
  (\bibinfo {year} {2005})}\BibitemShut {NoStop}%
\bibitem [{\citenamefont {Hainaut}\ \emph {et~al.}(2017)\citenamefont
  {Hainaut}, \citenamefont {Manai}, \citenamefont {Chicireanu}, \citenamefont
  {Cl{\'e}ment}, \citenamefont {Zemmouri}, \citenamefont {Garreau},
  \citenamefont {Szriftgiser}, \citenamefont {Lemari{\'e}}, \citenamefont
  {Cherroret},\ and\ \citenamefont
  {Delande}}]{Hainaut:EnhancedReturnOrigin:PRL17}%
  \BibitemOpen
  \bibfield  {author} {\bibinfo {author} {\bibfnamefont {C.}~\bibnamefont
  {Hainaut}}, \bibinfo {author} {\bibfnamefont {I.}~\bibnamefont {Manai}},
  \bibinfo {author} {\bibfnamefont {R.}~\bibnamefont {Chicireanu}}, \bibinfo
  {author} {\bibfnamefont {J.-F.}\ \bibnamefont {Cl{\'e}ment}}, \bibinfo
  {author} {\bibfnamefont {S.}~\bibnamefont {Zemmouri}}, \bibinfo {author}
  {\bibfnamefont {J.~C.}\ \bibnamefont {Garreau}}, \bibinfo {author}
  {\bibfnamefont {P.}~\bibnamefont {Szriftgiser}}, \bibinfo {author}
  {\bibfnamefont {G.}~\bibnamefont {Lemari{\'e}}}, \bibinfo {author}
  {\bibfnamefont {N.}~\bibnamefont {Cherroret}}, \ and\ \bibinfo {author}
  {\bibfnamefont {D.}~\bibnamefont {Delande}},\ }\bibfield  {title} {\enquote
  {\bibinfo {title} {{Return to the Origin as a Probe of Atomic Phase
  Coherence}},}\ }\href {\doibase 10.1103/PhysRevLett.118.184101} {\bibfield
  {journal} {\bibinfo  {journal} {Phys. Rev. Lett.}\ }\textbf {\bibinfo
  {volume} {118}},\ \bibinfo {pages} {184101} (\bibinfo {year}
  {2017})}\BibitemShut {NoStop}%
\bibitem [{\citenamefont {Hainaut}\ \emph
  {et~al.}(2018{\natexlab{c}})\citenamefont {Hainaut}, \citenamefont {Fang},
  \citenamefont {Rancon}, \citenamefont {Cl{\'e}ment}, \citenamefont
  {Szriftgiser}, \citenamefont {Garreau}, \citenamefont {Tian},\ and\
  \citenamefont {Chicireanu}}]{Hainaut:TimePhaseTrQuChaos:PRL18}%
  \BibitemOpen
  \bibfield  {author} {\bibinfo {author} {\bibfnamefont {C.}~\bibnamefont
  {Hainaut}}, \bibinfo {author} {\bibfnamefont {P.}~\bibnamefont {Fang}},
  \bibinfo {author} {\bibfnamefont {A.}~\bibnamefont {Rancon}}, \bibinfo
  {author} {\bibfnamefont {J.-F.}\ \bibnamefont {Cl{\'e}ment}}, \bibinfo
  {author} {\bibfnamefont {P.}~\bibnamefont {Szriftgiser}}, \bibinfo {author}
  {\bibfnamefont {J.~C.}\ \bibnamefont {Garreau}}, \bibinfo {author}
  {\bibfnamefont {C.}~\bibnamefont {Tian}}, \ and\ \bibinfo {author}
  {\bibfnamefont {R.}~\bibnamefont {Chicireanu}},\ }\bibfield  {title}
  {\enquote {\bibinfo {title} {{Experimental Observation of a Time-Driven Phase
  Transition in Quantum Chaos}},}\ }\href {\doibase
  10.1103/PhysRevLett.121.134101} {\bibfield  {journal} {\bibinfo  {journal}
  {Phys. Rev. Lett.}\ }\textbf {\bibinfo {volume} {121}},\ \bibinfo {pages}
  {134101} (\bibinfo {year} {2018}{\natexlab{c}})}\BibitemShut {NoStop}%
\end{thebibliography}

\end{document}